# Transport of Particles in Liquid Crystals

**Oleg D. Lavrentovich**

Liquid Crystal Institute and Chemical Physics Interdisciplinary Program,
Kent State University, Kent, OH 44240

***Abstract.*** *Colloidal particles in a liquid crystal (LC) behave very differently from their counterparts in isotropic fluids. Elastic nature of the orientational order and surface anchoring of the director cause long-range anisotropic interactions and lead to the phenomenon of levitation. The LC environment enables new mechanisms of particle transport that are reviewed in this work. Among them the motion of particles caused by gradients of the director, and effects in the electric field: backflow powered by director reorientations, dielectrophoresis in LC with varying dielectric permittivity and LC-enabled nonlinear electrophoresis with velocity that depends on the square of the applied electric field and can be directed differently from the field direction.*

## 1. INTRODUCTION

There is a growing interest to what can be called a soft active metamatter (SAMM), an emerging class of engineered composite soft materials[1], objects and systems with the ability to move and reconfigure structure and properties in response to external stimuli.[2-6] The grand challenge is to establish how to construct different forms of SAMM. The two important elements are (i) elementary moving units, or "meta-atoms", and (ii) the surrounding medium that controls them.

The concept of a meta-atom is similar to the one used in construction of optical metamaterials[7]. In the context of optics, a meta-atom is designed from dielectric and metallic elements to impart a specific type of local interaction with an electromagnetic wave, which results in unusual macroscopic features, such as a negative index of refraction. In the context of SAMM, the meta-atom can be pre-programmed to interact in a specific way with other units, to participate in spontaneous or directed self-assembly[8] and to carry a useful function, such as the ability to move. The role of the surrounding medium is to guide the meta-atoms in all these processes. Typically, self-assembly and especially its dynamic variations are considered to occur in an isotropic medium such as water. The trend starts to change, with the realization that the medium can enable useful properties that the meta-atoms would not display by themselves. Among them are active media with internal sources of energy (for example, a bacterial bath [9]), liquids with gradient properties [10], porous [11] and cellular [12, 13] structures, and anisotropic fluids. An electric [14] and magnetic field [15, 16] can also act as a functional medium, enabling anisotropic interactions and particle dynamics.

The purpose of this review is to discuss an elementary example of SAMM, a colloidal particle in a liquid crystal (LC). Replacement of an isotropic fluid environment with a LC leads to dramatic changes in static and dynamic behaviour of the colloidal particles (meta-atoms for the purpose of this review). Of especial interest are situations when the colloids are transported through (and, as we shall see, often thanks to) the LC medium, driven by the external electric field. First we provide a brief review of anisotropic properties of LCs, then describe statics of colloidal particles in them and, finally, review the recent explorations of colloidal transport in LCs.

## 2. BASIC PROPERTIES OF LIQUID CRYSTALS AND COLLOIDS

### 2.1. Orientational order and anisotropy of liquid crystals

Liquid crystals represent a state of matter with orientational order of molecules and complete (the case of nematics) or partial absence of the long-range positional order. In a uniaxial nematic, the direction of average orientation is specified by the director $\hat{\mathbf{n}}$ with the properties $\hat{\mathbf{n}}^2 = 1$ and $\hat{\mathbf{n}} \equiv -\hat{\mathbf{n}}$ [17-19]. The long-range orientational order makes all the properties of the nematic anisotropic. For example, permittivity $\varepsilon_\parallel$ measured with the electric field parallel to $\hat{\mathbf{n}}$ is generally different from the permittivity $\varepsilon_\perp$ measured in perpendicular direction. If the anisotropy $\varepsilon_a = \varepsilon_\parallel - \varepsilon_\perp$ is positive, the applied electric field $\mathbf{E}$ realigns $\hat{\mathbf{n}}$ parallel to itself; $\varepsilon_a < 0$ leads to perpendicular orientation. Similarly anisotropic are diffusion coefficients of guest molecules and ionic conductivities, $\mu_\parallel$ and $\mu_\perp$. There are five different viscosity coefficients.

Any deviations of the director from the uniform state cost some elastic energy; the corresponding free energy density is described by the Frank-Oseen functional featuring the so-called Frank elastic constants of splay ($K_1$), twist ($K_2$) and bend ($K_3$),

$$f_{elastic} = \tfrac{1}{2} K_1 \left(div\hat{\mathbf{n}}\right)^2 + \tfrac{1}{2} K_2 \left(\hat{\mathbf{n}} \cdot curl\hat{\mathbf{n}}\right)^2 + \tfrac{1}{2} K_3 \left(\hat{\mathbf{n}} \times curl\hat{\mathbf{n}}\right)^2 \quad (1)$$

to which one often adds the divergence terms such as the saddle-splay term [18].

### 2.2. Surface anchoring and two types of liquid crystal colloids

Anisotropic interactions at LC surfaces set one (or more) directions of preferred orientation $\hat{\mathbf{n}}_0$ of the director, called the easy axis. To deviate the actual director $\hat{\mathbf{n}}$ from $\hat{\mathbf{n}}_0$, one needs to spend some work. Typically, for deviations in the plane normal to the surface, the work is expressed by the so-called Rapini-Papoular potential,

$$f_{anch} = \tfrac{1}{2} W_a \sin^2\left(\theta - \theta_0\right), \quad (2)$$

where $W_a$ is the polar (out-of-plane) anchoring coefficient, $\theta_0$ is the angle between the easy axis and the normal to the interface; $\theta$ is the actual surface tilt of the director. Experimental data[20, 21] on $W_a$ vary broadly, $W_a \sim \left(10^{-6} \div 10^{-3}\right) \mathrm{J/m}^2$. For a typical thermotropic LC such as pentylcyanobiphenyl (5CB) aligned tangentially at a rubbed polyimide[22] (materials often used in preparation of electrooptical nematic cells), nylon [23], polyisoprene [24], or mica[25], the order of magnitude is $W_a \sim 10^{-4} \mathrm{J/m}^2$. Smaller values $W_a \sim \left(10^{-6} - 10^{-5}\right) \mathrm{J/m}^2$ are measured for normally anchored nematics[26], for example, at surfaces functionalized with surfactants[20]. For lyotropic LCs of chromonic type[27], $W_a \sim \left(10^{-6} - 10^{-5}\right) \mathrm{J/m}^2$. The surface anchoring strength and even the preferential alignment of the director can be influenced by ionic species in the system[26].

The ratio $\lambda_{dGK} = K / W_a$ of the average elastic constant to the anchoring coefficient has dimension of length and is called the de Gennes-Kleman length. For the typical $K = 10$ pN and $W_a \sim \left(10^{-6} \div 10^{-3}\right) \mathrm{J/m}^2$, one finds $\lambda_{dGK} = \left(0.1 \div 10\right) \mathrm{\mu m}$ to be much larger than the molecular scale $l \sim \left(1 \div 10\right) \mathrm{nm}$. The feature immediately leads to an interesting expectation regarding a colloidal particle in a LC.

Suppose a sphere of a radius $R$ favours a perpendicular (radial) surface anchoring. If it is placed in a uniform LC with $\hat{\mathbf{n}}(\mathbf{r}) = \mathrm{const}$, there are two principal ways to reconcile the two conflicting director fields. One can require the director to be uniform everywhere, $\hat{\mathbf{n}}(\mathbf{r}) = \mathrm{const}$. The energy penalty is the anchoring energy (integrated over the surface of the sphere) that scales with the area, $F_{anch} \sim W_a R^2$. The second solution is to satisfy the normal boundary conditions and distort the director in the vicinity of sphere so that is matches the uniform far-field. The elastic energy of these distortions, according to Eq.(1), scales mostly *linearly* with $R$:

$$F_{elastic} \sim K \left(\frac{\theta - \theta_0}{R}\right)^2 R^3 \sim KR. \quad (3)$$

The balance of two energies, $W_a R^2$ and $KR$ suggests that small particles, $R \ll K/W_a$, leave the surrounding director mostly uniform, Fig.1(a), while large particles, $R \gg K/W_a$, satisfy the surface anchoring and distort the director in the neighbourhood, Fig.1(b,c). These two extreme cases can be classified as liquid crystal colloids with weak intrinsic anchoring and strong intrinsic anchoring, respectively.

*(i) LC colloids with weak intrinsic anchoring.* If the initial director is uniform, it is likely to remain as such even if the small particles $R \ll \lambda_{dGK}$ are added. Although the director orientation does not change much, the small particles still impart new properties on the composition, through their contributions to the effective elastic, viscous, dielectric and oprical anisotropy, etc. LCs doped with ferroelectric and ferromagnetic submicron particles represent a system of this type [28-31]. The additive-modified molecular interactions can also change the overall director structure. A good example is an addition of chiral molecules, $R \sim l$ to a nematic that transform it into a cholesteric with a helicoidal director. There is also an opposite effect: the director imposes an orientational torque on small anisometric inclusions, such as dye molecules; this "guest-host" phenomenon is used in three-dimensional optical microscopy of the director fields (section 3.2).

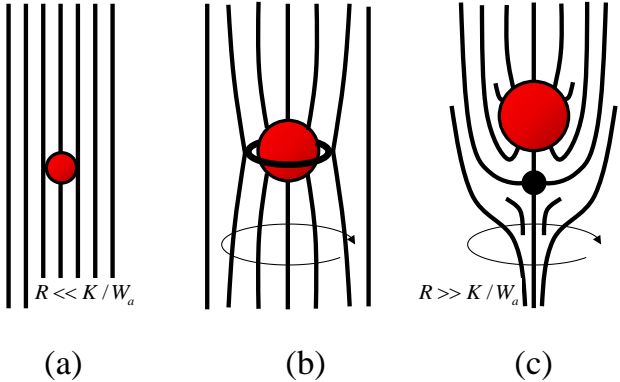

**Fig.1.** *Director fields around a sphere with perpendicular anchoring placed in a uniform nematic: (a) small particle, $R \ll K/W_a$, does not distort the director but violates boundary conditions; (b,c) large particles $R \gg K/W_a$ set director distortions in order to satisfy the surface anchoring conditions. Shown distorted configurations are of quadrupolar symmetry with an equatorial disclination loop (b) and of dipolar symmetry with a hyperbolic point defect-hedgehog (c).*

*(ii) LC colloids with strong intrinsic anchoring.*

The concrete type of distortions around a large $R \gg \lambda_{dGK}$ sphere depends on orientation of the easy axis. For perpendicular anchoring, the simplest geometry to imagine is the so-called Saturn ring structure [32] of quadrupolar symmetry, Fig.1(b). The matching between the local radial director $\hat{\mathbf{n}}_0 = \mathbf{r}/|\mathbf{r}|$ near the sphere and the uniform far-field is provided by a disclination ring of strength (-1/2). The Saturn ring structure, however, is usually unstable, as the disclination ring can shrink into a point defect, the so-called "hyperbolic" hedgehog [33], Fig.1(c), thus reducing the elastic energy. The resulting dipolar structure with a point defect-hedgehog is energetically stable in most cases of normal anchoring. The ring structure can be stabilized in a thin cell of thickness close to 2 $R$ [34], for relatively small $R$ or for relatively weak anchoring strength[35], under a high electric or magnetic field [36, 37]. Besides the configurations shown in Fig.1, more complex structures, including twisted ones, around particles of various shapes have been considered, see, for example, [38-43] and references therein. Figure 2 shows the optical microscopy textures of a sphere with normal anchoring and accompanying hyperbolic hedgehog and its modification by a strong alternating current (AC) electric field.

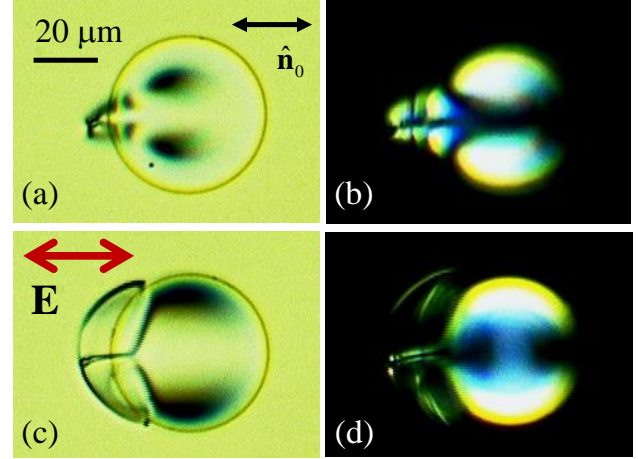

**Fig.2.** *Optical microscopy textures of a silica sphere with dipolar director distortions. Normal anchoring (surface functionalization with N,N-didecyl-N-methyl-(3-trimethoxysilylpropyl) ammonium chloride). Nematic LC mixture E7. Hyperbolic point defect on the left hand side of the sphere, viewed between two (a) parallel and (b) crossed polarizers; (c,d) expansion of the director deformations by a horizontal electric field $E = 0.3$ V/µm of frequency 1.25 kHz. Photos by Israel Lazo.*

The case of $R \gg K/W_a$ represents the most dramatic departure of the LC-colloid system from its isotropic counterparts, as the director distortions around the particles bring a number of new facets in the behaviour of colloids, such as anisotropic inter-particle forces[33], levitation[44] and nonlinear electrophoresis[45]. A rough estimate of the order of magnitude of the elastic energy associated with the director distortions around a strongly anchored micron-size particle placed in an otherwise uniform nematic cell, is $F_{elastic} \sim KR \sim (U_{LC}/l)R \sim k_B T_{NI}(R/l) \sim 10^3 k_B T_{NI}$, where one assumes for the elastic constant a de Gennes' value[17] $K \sim U_{LC}/l$, with $U_{LC} \sim k_B T_{NI}$ being the energy of molecular interactions responsible for the existence of the LC state at temperatures below the melting temperature $T_{NI}$ (close to the room temperature for most LCs); $k_B$ is the Boltzmann constant. These enormous (by colloidal standards) forces lead to fascinating examples of long-range anisotropic interactions, such as pairing of particles and formation of multi-particle chains and clusters [33, 39, 46-53]; see the recent reviews by Tkalec and Muševič [54], by Araki, Serra and Tanaka [55], and by Blanc et al[56]. Forces of the same nature are also responsible for attraction of colloidal particles to (particle-free) distortions and defects in the director field in nematics [47, 57, 58], smectics [59-61] and blue phases [62], for trapping and ordering of particles at the LC surfaces [63-66], and even for symmetry-breaking that enables transport phenomena such as nonlinear electrophoresis in LCs[45, 67]. As discussed in the next section, the balance of anchoring and elastic forces also leads to the effect of colloidal levitation in LC environment, thanks to the elastic repulsion from the bounding wall [44, 68-71] (an interesting version of the magnetic field-induced levitation in LCs has been presented by Lapointe et al [72, 73]).

Another important implication of the large energies of interactions of particles in LCs is that the structures assembled through the deformed director field might be prone to form metastable configurations with large energy barriers $\sim F_{elastic}$ separating them from other states, including the true equilibrium state. These considerations have been nicely illustrated in recent studies by Wood et al[74] who demonstrated that upon the increase in volume fraction $\phi$ of colloids in the nematic matrix, the system forms a defect-entangled gel with strongly increased elasticity and energy barriers $\sim F_{elastic} \sim (10^2 - 10^3) k_B T$ between different states.

### 2.3. Brownian motion of colloids in LCs

In its simple realization, Brownian motion is observed as random displacements of a small particle in an isotropic fluid, controlled by the kinetic energy dissipation [75]. The mean displacement is zero, but the average mean squared displacement (MSD) is finite, growing linearly with the time lag $t$ [76], $\langle \Delta \mathbf{r}^2(\tau) \rangle = 6Dt$, where $D$ is the translational diffusion coefficient. For a sphere in a fluid of viscosity $\eta$, according to the Stokes-Einstein relation, $D = k_B T / 6\pi \eta R$. In LCs, Brownian motion becomes anisotropic, with the coefficient $D_\parallel$ characterizing diffusion parallel to $\hat{\mathbf{n}}_0$ being different from the coefficient $D_\perp$ for perpendicular motion[67, 77-85],

$$D_{\parallel,\perp} = k_B T / 6\pi \eta_{\parallel,\perp} R, \quad (4)$$

where the viscosities $\eta_\parallel \neq \eta_\perp$ depend on molecular orientation around the particle, Fig.3.

The anisotropic character of diffusion in LCs has been established by Loudet et al.[80] for tangentially anchored spheres with quadrupolar director distortions. In the case of normal surface anchoring and dipolar distortions[67], presence of the topological defect near the sphere breaks the "fore-aft" symmetry, Fig.1c, 2a. Of course, this feature does not rectify Brownian diffusion of the spheres and does not result in unidirectional movement: the time-average displacement of the particles averages to zero, as clear from the probability distribution of particle displacements parallel and perpendicular to $\hat{\mathbf{n}}_0$, Fig. 3a. The dependence of parallel and perpendicular components of MSD on the time lag, Fig.3b, is linear and anisotropic, as in Eq.(4), with $D_\parallel / D_\perp = 1.7$.

Brownian particles in complex fluids may exhibit an anomalous behavior of MSD, $\langle \Delta \mathbf{r}^2(\tau) \rangle \propto t^\alpha$, with the exponent $\alpha$ either smaller than 1 (subdiffusion) or larger than 1 (superdiffusion). For example, subdiffusive motion of colloidal particles is observed in polymer [86] and F-actin networks [87], in surfactant dispersions [88]; superdiffusion occurs in concentrated suspensions of swimming bacteria [89, 90] and dispersions of polymer-like micelles [91-94]. The diffusion regimes reflect the properties of the host medium that might be ordered, inhomogeneous or contain internal energy sources; this connection forms the basis of microrheology. There are also reports of anomalous diffusion in orientationaly ordered systems with additional features, such as bacterial activity [89, 90], size distribution of building units [91], fluctuations of concentration [93], spatial modulation of hydrophobic and hydrophilic regions [92], bending rigidity of the molecular aggregates [92], etc. It is these additional factors that have been used to explain the anomalous diffusion. Recently, Pumpa and Cichos [95] noted that diffusion of dye molecules in LCs is much slower than self-diffusion; the effect was attributed to director distortions around the dye.

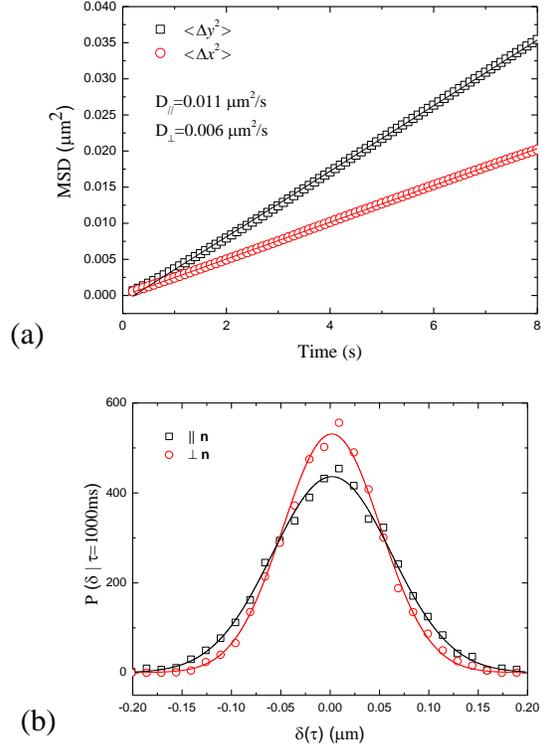

**Fig. 3.** *(a) Probability distribution of displacements of normally anchored silica spheres of diameter $2R = 5.1\,\mu m$ dispersed in E7, in the directions parallel and perpendicular to $\hat{\mathbf{n}}_0$ for time intervals of 1s. Solid lines represent Gaussian fits. (b) MSD vs time lag for the two components of displacement; solid lines represent linear fits.* $25^0 C$. *I. Lazo's data, Ref.[67].*

In Fig.3, the time lags are on the order of 0.1 s and larger. These time scales are larger than the typical time scale of director relaxation around the sphere, which can be estimated as[17] $\tau \sim \beta^2 \eta_{\parallel,\perp} R^2 / K$, where $\beta$ is a numerical coefficient of the order of 1 that describes the length scale $\sim \beta R$ of director deformations around the particle. For a typical nematic LC, such as 5CB or E7, $\eta_{\parallel,\perp} / K \sim 10^{10}$ s/m$^2$ and the relaxation time for a micron-sized particles would be of the order of $\tau \sim (0.01 - 0.1)\,s$. At these time scales, one can expect that the particle diffusion will be strongly affected by the director deformations and their relaxation. The LCs are viscoelastic media, in which the director field $\hat{\mathbf{n}}(\mathbf{r},t)$ is coupled to the velocity field $\mathbf{v}(\mathbf{r},t)$. Both $\hat{\mathbf{n}}(\mathbf{r},t)$ and $\mathbf{v}(\mathbf{r},t)$ are perturbed by the particle and by the director fluctuations. Translational motion is coupled to the orientational dynamics of $\hat{\mathbf{n}}(\mathbf{r},t)$. In its turn, director reorientations induce torques and forces that cause the material to flow (the so-called backflow effect) and thus modify $\mathbf{v}(\mathbf{r},t)$.

## 3. LEVITATION OF PARTICLES IN BOUNDED LIQUID CRYSTAL

### 3.1. Sedimentation

If the particle density is different from density of the surrounding fluid, $\Delta\rho = \rho_p - \rho$, gravity causes sedimentation. For $\Delta\rho > 0$, the particle drops to the bottom of container. Sedimentation can be opposed by thermal (Brownian) motion. To compare the relative importance of the two factors, consider the barometric formula that describes the probability of finding a sphere with an excess mass $m^* = \frac{4}{3}\pi R^3 \Delta\rho$, at some height $z$ above the bottom of the container[96], $p(z) = \exp(-m^* g z / k_B T)$, where $g \approx 9.8$ m/s$^2$ is the standard gravity. The thermal motion can raise the sphere up to the height $z_s = \frac{3k_B T}{4\pi R^3 g \Delta\rho}$, called the sedimentation (or gravitation) length. The sedimentation length is used to define the term "colloid", which is applicable to a particle smaller than $z_s$. For typical materials such as glass, polymers, dispersed in water at room temperature, with an excess density $\Delta\rho \sim 10^2$ kg/m$^3$, $z_s \approx 1$ μm and the definition of colloid is satisfied as long as the particle is of a submicron size [96, 97]. In a LC, elastic levitation can keep the particle at a distance above the container's bottom that is much larger than the sedimentation length.

### 3.2. Liquid-crystal-enabled levitation

In a LC sample, sedimentation is opposed not only by thermal motion, but also by elastic repulsion from bounding walls [44], Fig.4. Consider a sphere with a radial anchoring and an accompanying hyperbolic hedgehog, placed in a semi-infinite nematic bulk, bounded by a rigid flat wall at $z=0$ with a strong surface anchoring parallel to the wall. The sphere is repelled from the wall, as the uniform director field near the wall is incompatible with the director distortions around the sphere. The elastic potential of repulsion, in the dipole approximation, neglecting in-plane anchoring effects, writes as [44]:

$$F_{repulsion} \approx A^2 \pi K \frac{R^4}{z^3}, \quad (3)$$

where the dimensionless coefficient $A$ (of the order of few units) depends on factors such as surface anchoring strength, elastic anisotropy of the material, etc. [33, 69, 98]. Here again, the elastic energy is much larger than the thermal energy: $F_{repulsion} \approx 60 k_B T$ for $A = 1$, $K = 10$ pN, $R = 1$ μm, $z = 5$ μm. When the elastic repulsion from the wall competes with gravity, the equilibrium height of the elastically levitating particle is $z_{elastic} = R(3\pi A^2 K / m^* g)^{1/4}$, or

$$z_{elastic} = \left(\frac{3}{2}A\right)^{1/2} \left(\frac{KR}{\Delta\rho g}\right)^{1/4}. \quad (4)$$

For $A=1$, $K=10$ pN, $R=1$ μm, $\Delta\rho \sim 10^2$ kg/m$^3$, one finds $z_{elastic} \sim 10$ μm, one order of magnitude higher than the sedimentation distance $z_s \sim 1$ μm. The estimate is confirmed by the experiment [44], in which one determines the shift $\delta$ of the sphere's center from the mid-plane $z = h/2$ of a cell formed by two parallel glass plates; $z_{elastic} = h/2 - \delta \sim 10$ μm in Fig. 4b,c. The overall dependence $\delta(h)$ is described well by the balance of gravity and two elastic forces [44], Fig. 4c.

The levitating particles can be seen in the images of the vertical cross-sections of the cells, made possible by the confocal microscope, Fig.4(b). A modification of a standard confocal microscopy, the so-called fluorescence confocal polarizing microscopy (FCPM), allows one also to obtain images of the 3D director field[99]. The textures shown in Fig.4 (b) represent the vertical cross-sectional views of the "sandwich" nematic cell, bounded by two flat glass plates at the bottom and top (compare to regular polarizing microscopy textures in Fig.2 that show the "horizontal" image of the sample). The FCPM mode of observation is enabled when the LC is doped with dye molecules of elongated shape that are aligned by the director. The sample is scanned with a focused polarized laser beam. The intensity of fluorescent signal depends on the angle between the direction of light polarization and $\hat{\mathbf{n}}$ (which is also the direction of transition dipole of the dye molecules). The intensity is maximum when the polarization and $\hat{\mathbf{n}}$ are parallel and minimum when the two are mutually perpendicular. FCPM allows one to obtain 3D images of complex director structures. The colloidal particles used in this work are not fluorescent and appear as dark regions in the textures.

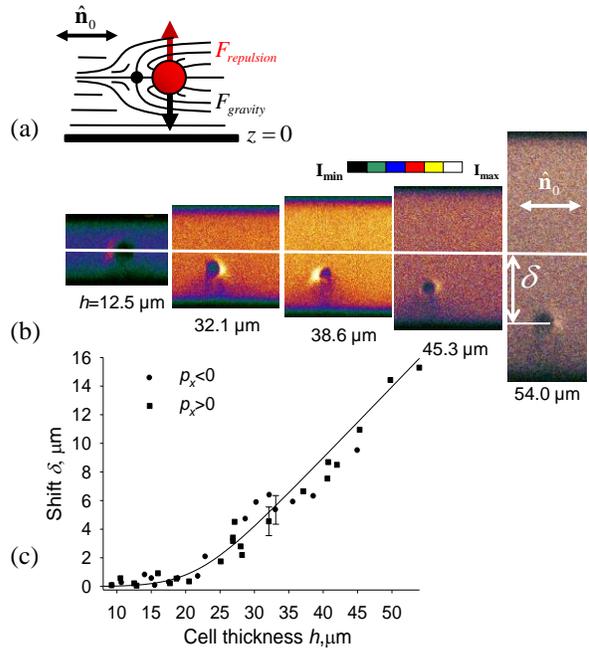

**Fig.4.** *(a) Balance of gravity and elastic force keeps the sphere levitating in the nematic bulk of a planar cell; (b) FCPM textures of vertical cross-sections of sandwich cells of different thickness $h$, showing levitating silica spheres; the bright spot next to the dark sphere is the hyperbolic hedgehog; (c) shift $\delta$ of the particle's center from the mid-plane $z = h/2$ vs. cell thickness $h$, see Ref.[44] and text for more details.*

The balance of elastic and gravity forces, Eq.(4), predicts that $z_{elastic} \propto R^{1/4}$, i.e. the height at which the particle levitates *increases* with the particle's size, an effect opposite to the dependence $z_s \propto 1/R^3$ for the isotropic medium. The increase of $z_{elastic}$ with $R$ has been demonstrated experimentally for spheres with normal anchoring [68], Fig.5.

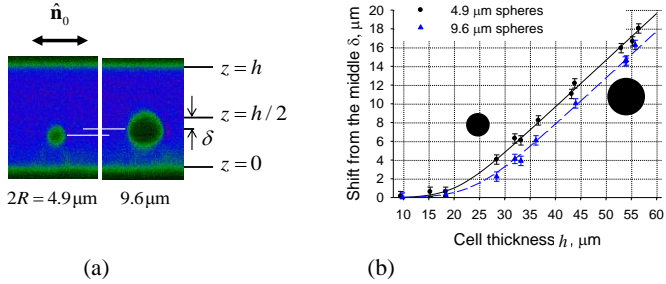

(a)    (b)

**Fig.5.** *FCPM cross-sectional view of the nematic cell demonstrating that the larger particle (a) levitates at the higher level than the smaller particle (b). Shift of the spherical particles of diameter 4.9 μm (a) and 9.6 μm (b) from the middle plane of the nematic cell vs. cell thickness [68] (c).*

Sedimentation is detrimental to 3D assembly in isotropic fluids. Self-assembly of ordered structures, although offering a broad technological platform [8], often requires a guiding support of quasi 2D templates (substrates) [100, 101]. The development of practical strategies for the assembly of 3D arrays remains an unsolved problem [102]. One of the most promising approaches is magnetic levitation [102] that allows one to assemble 3D arrays of diamagnetic particles in a paramagnetic fluid medium by using a gradient magnetic field. Elastic levitation in a LC medium offers another useful approach to 3D assembly. This specific LC-enabled effect can be combined with other means to control location and transport of particles and to resist the forces of gravity. Below we consider one of the simplest and universal effect, linear electrophoresis.

### 3.3. Linear electrophoresis.

Electrophoresis is a motion of an electrically charged particle relative to a fluid in a uniform electric field [103], despite the fact that the system is neutral as a whole, since the charge of particle is compensated by counterions in the medium. In an isotropic fluid, the electrophoretic velocity of the particle is linearly proportional to the applied field, $\mathbf{v} = \mu_l \mathbf{E}$, where the electrophoretic mobility $\mu_l$ is proportional to the particle's charge and inversely proportional to the fluid's viscosity.

A similar effect can be staged in a LC. As an example, consider a planar cell $\hat{\mathbf{n}}_0 = (1,0,0)$ filled with a nematic of negative dielectric anisotropy, $\varepsilon_a < 0$. The field is vertical, $\mathbf{E} = (0,0,E_z)$, so that it does not perturb the overall director $\hat{\mathbf{n}}_0$. If the particle is charged, it will be moved by the field along the vertical $z$-axis, Fig.6. The direction depends on the sign of particle's charge and the field polarity.

The $z$ position of the levitating particle is determined by the electrophoretic force, gravity, elastic repulsion from the walls, etc. The electrophoretic force diminishes with time because moving ions of both signs, always present in LCs, eventually block the electrodes. This is one of disadvantages of the standard DC-driven electrophoresis with a linear response to the applied electric field $v \propto E$. As we shall see in Section 8, the LCs enable a principally different mechanism of electrophoresis [45, 67] with a quadratic dependence $v \propto E^2$ that eliminates many of the drawbacks of the linear electrophoresis by allowing AC driving.

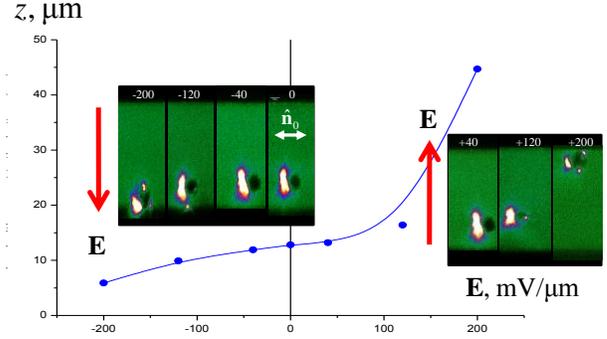

**Fig.6.** *$z$-Positions of electrophoretically controlled levitation[67] of borosilicate glass spheres of diameter $2R = 9.6$ μm in the nematic cell visualized by FCPM [67]; the spheres move down when $E_z < 0$ and up when $E_z > 0$.*

## 4. TRANSPORT OF PARTICLES BY GRADIENTS OF THE ORDER PARAMETER

### 4.1. Static distortions

LC-enabled levitation of colloids can adopt a variety of forms. For example, the cores of line defects that frame focal conic domains in smectic LC can lift spherical particles above the base of the domain to minimize the surface anchoring energy [59]. Colloidal particles have been shown to be attracted by the cores of point defects [33] and disclination lines [58, 104]. An example of accumulation of polymer particles at the core of a singular disclination of strength 1/2 is illustrated in Fig. 7(a). Qualitatively, the effect is understood as following. In a uniform nematic, Fig.7(b), there is no preferred location for the particles. If the director is distorted over the scale $\xi$, the particle is generally attracted to the region of maximum distortions, as it replaces the energetically costly director field with itself. The gain in the elastic energy is roughly $\Delta F_{elastic} \sim -\frac{K}{\xi^2}R^3$, Fig.7(c), which can be easily much larger than $k_B T$; this is the case, for example, when a relatively small particle of size $R \sim (10-100)$ nm replaces strong distortions with $\xi \sim (10-100)$ nm. Of course, other factors, such as modified surface anchoring and entropy driven randomization should also be taken into account, as discussed in Refs. [58, 105] and in the recent review by Blanc et al[56]. Entropic forces become especially important when the typical size of the inclusions in in the range of nanometers. Here one can still expect that the nanoparticles would be attracted to the regions where the order is strongly distorted (as demonstrated, for example for 5 nm particles forming chains along the linear defects in smectic films [106]). Another effect at nanoscales is that the anisometric particles are aligned by the surrounding director, as already discussed.

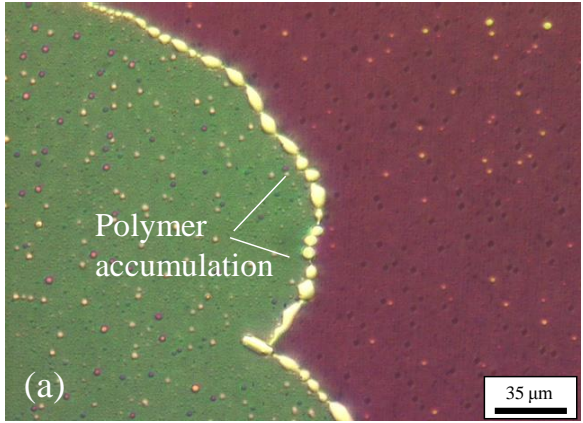

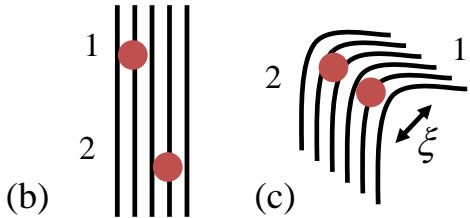

**Fig.7.** *(a) Accumulation of polymer at the disclination line separating two twisted nematic domains of opposite handedness. The monomer is dispersed in the nematic and photopolymerized by UV irradiation. Although the polymerized particles appear everywhere in the cell, they are driven to the disclination core where they replace the energetically costly strongly distorted director field. See Ref. [58] for more details. The schemes (b,c) explain the effect: the distorted regions reduce their energy by replacing the distorted director with the particles.*

As follows from the consideration above, location of a particle in a LC is dictated not only by gravity and Brownian motion, as in an isotropic fluid, but also by the director field, its gradients, balance of elastic and surface anchoring forces, as in Fig.7, by geometry and properties of bounding walls and interfaces, Fig.4,5 [63, 64], by presence of other particles [33], etc. Spectacular effects of linking colloidal particles into clusters by disclination lines have been demonstrated by Ljubljana group[49, 54]. By applying the electric or magnetic field to the LC, one can further diversify the set of parameters to control the topological defects and particles. One of the recent examples is manipulation of topological states filling the voids of a porous matrix thus creating different memory states [55, 107, 108]. Last, but not least, the particles can be driven by director distortions induced by focused laser beams [109, 110]. Unlike the case of regular optical tweezers trapping high-refractive-index particles in an isotropic medium with a low refractive index, optical tweezers in a LC can manipulate particles with arbitrary refractive index [111].

The location of particles in a nematic LC can be controlled not only by the gradients of the director field but also by the gradients of the scalar order parameter, describing the degree of molecular order [112, 113]. The local value of the scalar order parameter can be modified in a variety of ways, for example, by photo-induced conformational switches (say, trans-cis isomerization of azobenzene molecules doping the LC [112]) or simply by local heating of the LC [113, 114] with a laser beam. It turns out that the micrometer-scale particles are attracted to the regions of the reduced order parameter [113, 114]. Moreover, similar accumulation is observed at the molecular level [112, 113], for fluorescent molecules. The latter effect disappears in the isotropic phase of the LC[113].

Periodic variation of materials density in smectic LCs offers another mechanism of particle's transport. Imagine a smectic A LC doped with azobenzene derivatives. Elongated trans- isomers would prefer to locate themselves parallel to the host molecules forming the smectic layers, while the irregularly shaped cis-isomers would show affinity to the disordered interlayer regions [115]. By photo-addressing the sample and changing the trans-to-cis ratio, one can drive the azobenzene molecules into and out of the smectic layers [116, 117]. Relocation of the molecules changes the smectic layer spacing [118] which in its turn causes mechanical instabilities such as Helfrich-Hurault undulations of layers [119].

Recent studies of molecular transport in smectic LCs revealed a number of interesting features. Experimental and numerical studies of lyotropic smectics formed by long rigid rods (such as filamentous fd virus) demonstrated a hopping transport[120, 121]: the rods jump from one layer to another without significant reorientation. However, when the smectic is formed by flexible molecules, the layer-to-layer translocation can also happen through an intermediate state, in which the molecules find themselves between the layers in a reoriented state[122].

Control of particles' placement implies the ability to transport the particles, by designing the local properties of medium. In the following subsection 4.2, we consider transport of particles made possible by dielectrically created director gradients in a nematic cell.

**4.2. Elasticity-mediated transport in dielectrically addressed liquid crystal**

The LC-enabled levitation, Figs.4-6, opens new possibilities for the tasks that involve transportation and delivery, as it keeps the particles away from the bounding surfaces, thus reducing detrimental effects such as surface trapping and stronger Stokes drag near the boundaries.

We characterize the director configuration around a sphere with normal anchoring by an elastic dipole $\mathbf{p}=(p_x,0,0)$ directed from the hyperbolic hedgehog towards the sphere, Fig.8. It is either parallel to the $x$ axis ($p_x > 0$) or antiparallel to it ($p_x < 0$). The two states are separated by a large energy barrier. The sign of $p_x$ has no impact on the $z$-location, Eq.(4), as long as the overall director is horizontal, $\hat{\mathbf{n}}_0=(1,0,0)$. The situation changes if the LC is distorted by an applied electric field.

We consider a nematic of positive dielectric anisotropy, $\varepsilon_a > 0$. The electric field applied across the cell, Fig.8, realigns $\hat{\mathbf{n}}$ along the vertical axis $z$, but mainly in the center of the cell, $z \approx h/2$. Near the bounding plates, the director is fixed by the surface anchoring. The sub-surface regions with the strongest director distortions attract the spheres, lifting them from the middle plane, Fig.8(b). There are at least two reasons for the particles lift, elastic trapping [58] and LC-enabled dielectrophoresis [123, 124]. The first mechanism is discussed below, the second in Section 5.2.

The spatial scale of director distortions in an electric field can be quantified by the field-dependent dielectric length $\xi_E = \dfrac{1}{E}\sqrt{\dfrac{K}{\varepsilon_0 \varepsilon_a}}$ that is the distance over which the LC changes orientation from the easy axis to the direction along the field. The particle replaces the distorted LC region with itself. The associated energy reduction is [58]

$\sim KR^3/\xi_E^2$ for $R < \xi_E$ and $\sim KR$ for $R > \xi_E$ [123]. In a strong field ($E = 0.5\ V/\mu m$, $\xi_E \approx 0.7\ \mu m$), the elastic lifting force near the substrate is $\sim KR/\xi_E \approx 40$ pN, if one uses the typical estimates for a nematic such as 5CB or E7. Because of the opposite polarity of splay deformation $\text{div}\,\hat{\mathbf{n}}$ in the top and bottom parts of the cell, the particles with $p_x > 0$ and $p_x < 0$, initially at $z_{elastic} \approx h/2$, move towards the opposite plates. The initial population of identical spheres splits into two subsets, within each of which the dipoles are oriented in the same direction.

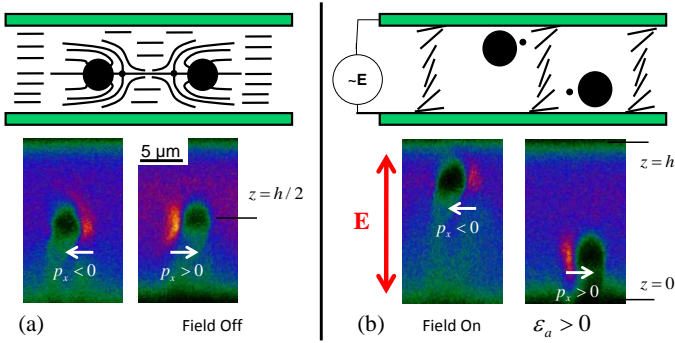

**Fig.8.** *(a) Two colloidal spheres with opposite elastic dipoles are located at the same height in a planar nematic cell, as seen in the FCPM textures of the vertical cross-sections of the cell; (b) Applied electric field reorients the director along the vertical direction; particles with opposite direction of dipoles move towards opposite plates as dictated by the polarity of splay deformations produced by the electric field*[44].

## 5. DIELECTROPHORESIS IN LCs

### 5.1. Dielectrophoresis in a non-uniform electric field

By definition, dielecrophoresis is a motion of matter caused by electric polarization in a non-uniform electric field[125]. Typically, the gradients of the electric field are created by a special geometry of electrodes. For example, Pohl[125] considered a coaxial electrode system that produces radial electric field. Imagine a particle in a space between the electrodes. An applied electric field polarizes the particle, separating the internal electric charges. Although the charges at the two ends of the particles are the same, the charge in a stronger electric field experiences a stronger electrostatic force. As a result, a polarizable sphere moves towards the maximum of the field, i.e. to the central electrode. Reversal of the field polarity does not change the result. The dielectrophoretic force $\mathbf{f}_d \propto (\varepsilon_p - \varepsilon_m)\nabla \mathbf{E}^2$ is independent of the field polarity; here $\varepsilon_p$ and $\varepsilon_m$ are the dielectric permittivities of the particle and the medium between the electrodes (in our example, $\varepsilon_p > \varepsilon_m$).

The effect can be used to drive particles of various shapes. In the case of a dispersion of metallic rods, the driving force not only concentrates the particles near the central electrode but also aligns them, which results in a significant field-induced orientational order and birefringence[126, 127].

Dielectriphoresis is used to transport solid particles and also to spread and transport isotropic fluids [128]. In the context of LC, it was first used to phase separate a LC and a photo-polymerizable monomer [129]. An electrode system represented square pixels of indium tin oxide (ITO) with non-conductive gaps separating them. In the vertical electric field, the LC, as a more polarizable component, moves into the space between the electrodes, while the monomer gathers in the gaps where it polymerizes. The electric field gradients can also be created with an in-plane electrode system[130]. The dielectriphoretic mechanism can be used to transport gold nanoparticles in the LCs[131].

Non-local character of dielectric effects in LCs enables a new type of dielectrophoresis, in which the gradients of the electric field and the ensuing dielectrophoretic force $\mathbf{f}_d \propto \nabla \mathbf{E}^2$ are caused by a varying dielectric permittivity of a LC with a distorted director, rather than by a special geometry of the electrodes[124], as discussed in the next subsection.

### 5.2. LC-enabled dielectrophoretic in a non-uniform director field

To illustrate the principle, we return to Fig.8, in which the nematic LC is confined between two perfectly flat and parallel electrodes. Above the splay Frederiks threshold, the applied electric field causes director reorientation. The dielectric permittivity in the distorted nematic changes from $\varepsilon(z = h/2) \approx \varepsilon_{//}$ in the middle of the cell, where $\hat{\mathbf{n}}$ is vertical, to $\approx \varepsilon_{\perp}$ near the boundaries, where the director is horizontal, $\hat{\mathbf{n}} \approx \hat{\mathbf{n}}_0$. A dielectric sphere in the gradient dielectric permittivity experiences a force resolved along the $z$ axis [123], moving to the regions with a lower $\varepsilon$, which in our case are the boundary regions. In a distorted LC, the dielectrophoretic force occurs even if the electrodes are flat and parallel, as long as $\varepsilon$ varies in space [68, 123]. In our case, $\varepsilon(z) = \varepsilon_{\perp}\sin^2\theta(z) + \varepsilon_{||}\cos^2\theta(z)$, where $\theta(z)$ is the angle between $\hat{\mathbf{n}}$ and the vertical field $\mathbf{E}$. The dielectric force is typically of the same order of magnitude as the elastic forces considered earlier [123, 124].

## 6. BACKFLOW-INDUCED BIDIRECTIONAL MOTION AND AGGREGATION OF COLLOIDS

In this section, we consider the electrically driven transport of particles controlled by backflow, resulting from the field-induced realignment of the director in LC cells. Such an effect is simply impossible in the isotropic liquids. The electric charge of the particle is irrelevant in this mechanism (although the charges, if present, can contribute to the dynamics through electrophoresis). The examples illustrate that in a LC, the external electric or magnetic field couples to a colloidal particle much more strongly than in isotropic liquids, thanks to existence of the director field.

### 6.1. Electric field induced bidirectional transport

Reorientation of the director in a bounded LC cell generates a backflow. The backflow, in its turn, affects the director reorientation [132, 133]. Zou and Clark [134] used a ferroelectric smectic C to show that the backflow can be used for unidirectional mass pumping of the LC. Small particles added to the LC, are also carried by backflow, as demonstrated clearly in the experiments on twisted nematic cells [135]. Below we consider an electrically driven planar nematic cell, in which the backflow effect results in bidirectional transport of colloids [44].

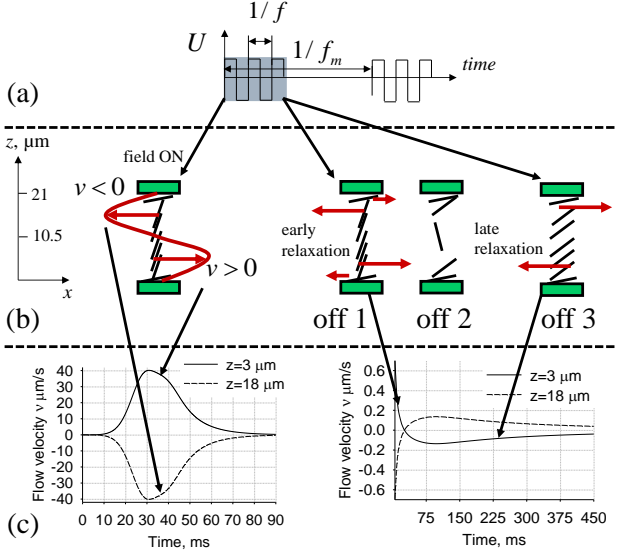

**Fig.9.** *Modulated electric field drives backflow in a planar nematic cell with $\varepsilon_a > 0$. (a) The cell is driven by periodic rectangular AC voltage pulses with a carrier frequency $f$ and a modulation frequency $f_m$, to impose a dielectric reorienting torque on the director. (b) The field on and field off switching causes director reorientation and material flows, see text. (c) Evolution of horizontal component of backflow velocity at two locations, $z=3$ μm and $z=18$ μm, in the cell of thickness 21 μm, simulated numerically in Ref. [44]; note the difference between the field-on and field-off regimes.*

As discussed above, the electric field separates the spheres with $p_x > 0$ and $p_x < 0$, moving them to the opposite plates as dictated by symmetry of the director and by the polar character of distortions around the particles, Fig.8(b). Once the particles are near the walls, they are caught into two anti-parallel backflow streams, Fig.9(b). The velocity v of the nematic is zero at $z = 0; h$ (when the no-slip condition is fulfilled, see [136] for the discussion of finite slips) and also in the middle plane. The antisymmetric profile of $\mathbf{v}(z)$ can be calculated numerically using the Ericksen-Leslie model, Fig.9(c), under an assumption that $\hat{n}$ is restricted to the vertical $xz$ plane and that there is only a horizontal component of velocity, $\mathbf{v} = [v(z,t), 0, 0]$ that depends on the $z$-coordinate and time $t$; the spheres and director distortions around them are excluded from the numerical simulations. The flow inertia can be neglected, as the flow relaxation time on the order of 1 μs is much shorter than the director relaxation times, $\tau_{on} \sim (\alpha_3 - \alpha_2)h^2 / \varepsilon_0 \varepsilon_a U^2 \sim$ 10 ms for the voltage $U$ switched on, and $\tau_{off} \sim (\alpha_3 - \alpha_2)h^2 / \pi^2 K_1 \sim$ 1 s for the voltage switched off; $\alpha_2$ and $\alpha_3$ are the nematic viscosities. Numerical simulations of $v(z,t)$ were performed using the material parameters characteristic for E7, typical $U = 10$ V and $h = 21$ μm, the same as in the laboratory experiments [44, 68]. The simulations reveal the following features, Fig.9(b,c).

When the field of carrier frequency $f$ (sufficiently high to avoid electrohydrodynamics caused by ions) is applied to the cell, Fig.9(a), $\hat{n}$ reorients in the center, say, counter-clockwise. The reorientation causes backflow with the velocity $v$ along the axis $x$ in the lower half of the cell and in the opposite direction in the upper half, Fig.9(b). Time evolution of $v$ at $z=3$ μm and $z=18$ μm is illustrated in Fig.9(c). When the voltage is switched on, the backflow velocity quickly reaches a maximum of about 40 μm/s and then fades to 0. Dynamics is very different when the field is switched off, right hand side of Fig.9. At the early stage, the director relaxation (clockwise rotation) starts near the plates, since in there the elastic distortions are the strongest. The coupling of reorientation to flow causes two local sets of anti-symmetric streams with velocities shown in red in Fig.9(b), the stage labelled "off 1". This flow, in its turn, creates a counterclockwise torque in the central portion of the cell; $\hat{n}$ can even flip over by more than 90 degrees from its original horizontal state, see stage "off 2" in Fig.9(c). This early stage is followed by a "proper" clockwise reorientation of $\hat{n}$ in the center and by a reversal of flow (stage "off 3"; note the sign change of velocity at around 35 ms in Fig.9(c)).

When the nematic is driven by a sequence of voltage pulses, modulated with a frequency $f_m \ll f$, the complex dynamics, Fig.9, results in net propulsion of particles, with the average velocity $v_p \approx \langle v \rangle$ close to the averaged velocity $\langle v \rangle$ of the fluid, Fig.10. The particles near the top move towards $+x$ and the particles near the bottom move towards $-x$. The velocity dependence on the modulation frequency is non-monotonous, with a pronounced maximum at $f_m^* \sim 1/(\tau_{on} + \tau_{off})$, which is easy to understand from the following qualitative consideration[44].

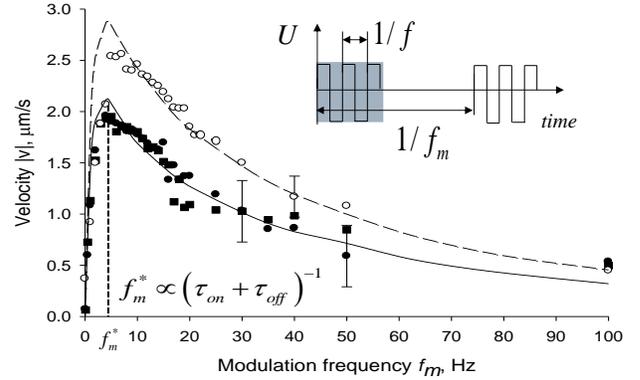

**Fig.10.** *Velocity of particles carried by backflow as a function of modulation frequency of the electric field; experimental data are taken from Ref. [44]. The maximum is achieved at the frequency determined by the director relaxation time.*

At low frequencies, $f_m \ll f_m^*$, the director has enough time to equilibrate and the velocity increases linearly with the frequency ( $f_m$ determines how many times the particles are pushed by the flow per unit time). At high frequencies, $f_m \gg f_m^*$, the director has no time to equilibrate, so that eventually, at $f_m \to \infty$, the motion stops. The maximum velocity is achieved at $f_m^* \sim 1/(\tau_{on} + \tau_{off})$.

An important issue in the study of colloidal dynamics in LC is whether the motion of the particle can modify the director field around it. The answer depends on the Ericksen number, which is the ratio of the elastic $\sim K/R^2$ and viscous $\sim \eta v / R$ torques:

$$Er = \frac{\eta R v}{K}.$$

Here $\sim \eta v / R$ is a characteristic LC viscosity involved. With the typical $\eta \sim 0.5\,\text{Pa}\cdot\text{s}$, $K \sim 10\,\text{pN}$, $R \sim 2\,\mu\text{m}$, one arrives at a conclusion that the Ericksen number exceeds 1 for velocities $v \sim 10\,\mu\text{m/s}$ and higher; these velocities are expected to influence the director configuration around colloids[137-139]. As follows from Fig.9(c), these velocities can be easily achieved in dielectric reorientation of the nematics.

We recall now that the polarity of splay deformations in the electric field separates the colloidal particles with opposite elastic dipoles, lifting them to the opposite plates, Fig.8(b). Backflow involves them in a bidirectional flow, so that the particles of one polarity move against the particles of the opposite polarity. The cell can be used as a "nematic collider" in which the particles with long-range anisotropic interactions inelastically collide and aggregate. Below, we describe how one can control the geometry of ensuing aggregates in such a collider[123, 124].

### 6.2. Backflow-induced aggregation in a nematic collider

Typically, in an isotropic dispersive medium the colloidal particles interact through central forces that do not depend on direction in space. When the interactions are anisotropic, the system becomes much harder to study but also more promising in the design of new materials [140], as demonstrated by studies of magnetic dipolar colloids [15, 141-143], Janus particles [144-146], colloids with electric [147-151] and magnetic [152, 153] field-induced dipoles. Anisotropy of interactions is reflected in aggregation geometry. For example, diffusion limited aggregation of spheres with central forces produces structures with fractal dimension $d_f$ between 1 and 2 in two dimensions (2D) and between 1 and 3 in 3D. Once dipolar interactions are switched on, $d_f$ reduces dramatically, down to 1 in both 2D and 3D, reflecting head-to-tail chaining [141, 152, 153]. For isotropic dispersive media, manipulating the particles shape has been the most effective way of tailoring the anisotropy of their interactions, as reviewed by Sacanna et al[154].

LCs represent a unique dispersive environment in which the anisotropy of colloidal interactions is created by the host medium itself. These interactions result in anisotropic aggregation. In the nematic collider, Fig.11, the overall geometry of aggregates is controlled by the distance (impact parameter) $b$ between the particles with $p_x > 0$ and $p_x < 0$ moving in two different planes. We label the two possible directions of the dipole as ">" and "<".

When there is no electric field, the colloidal particles are distributed randomly at approximately the same $z$ coordinate, Fig.8(a). One can use optical tweezers to bring them one-by-one into clusters stabilized by anisotropic interactions [46, 48, 49]. The second approach is to use collisions in a markedly non-equilibrium process, with an energy influx from the voltage pulses that cause reorientation and backflow.

In the bidirectional flow, Fig.11(a),(b), the particles can aggregate with neighbours, which are either of the same elastic polarity, moving near the same bounding plate, or of opposite polarity, moving near the opposite plate in the opposite direction. In the first case, the geometry of collision is ">>" and "<<", Fig.11(c); in the second case, it is of the type "$\genfrac{}{}{0pt}{}{>}{<}$", Fig.11(d). The prevalence of the two scenarios depends mostly on the impact parameter $b$. A large $b$ results in end-to-head attraction of colloids and formation of chains parallel to $\hat{\mathbf{n}}_0$, Fig.11(c). A small $b$ results in side-by-side attraction and formation of zig-zag chains perpendicular to $\hat{\mathbf{n}}_0$, Fig.11(d).

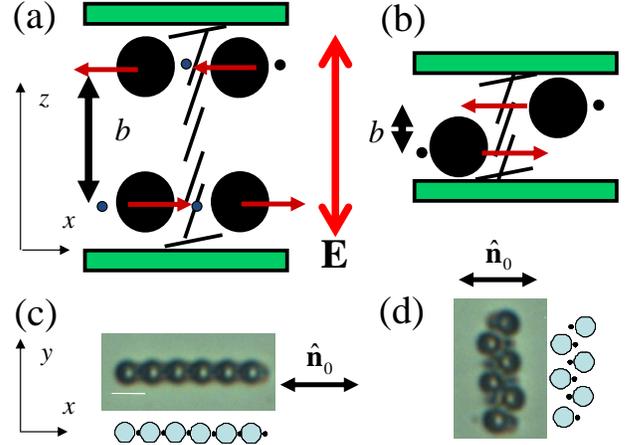

**Fig.11.** *Scheme of a nematic collider with (a) large and (b) small impact parameter $b$, facilitating chain aggregation (c) along the overall director and (d) perpendicular to it, respectively.*

A real aggregate grows with both scenarios involved. The analysis of relationship between the cluster's area (number $N$ of pixels in the image) and diameter $D$ reveals that clusters exhibit fractal behavior $N \propto D^{d_f}$, Fig.12 [123]. The fractal dimension $d_f$ depends on the cell thickness, Fig.12. Thicker cells yield smaller $d_f$ because the colloids collide predominantly head-to-tail with the neighbours of the same polarity and thus form aggregates through linear chaining along $\hat{\mathbf{n}}_0$, Fig.11(a,c), 12(a).

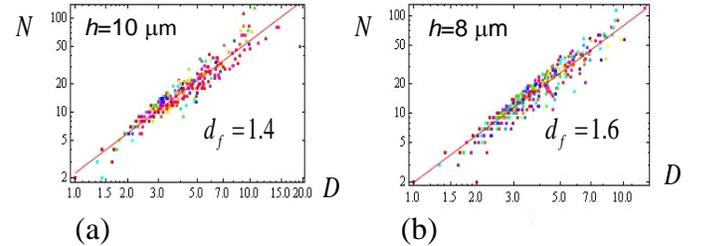

**Fig.12.** *Log-log plot of number of pixels $N$ and cluster diameter $D$ revealing the clusters fractal behavior $N \propto D^{d_f}$ in cells of thickness $10\,\mu\text{m}$ (a) and $8\,\mu\text{m}$ (b) at different time intervals. Time dependence of fractal dimension $d_f$ (filled markers) and anisotropy $q$ (empty markers) of anisotropic aggregates in $10\,\mu\text{m}$ (squares) and $8\,\mu\text{m}$ (circles) cells [124].*

In the described experiment, the building unit, or "meta-atom", interacting anisotropically with other "meta-atoms", is an elastic dipole. The dipolar model is only a rough approximation, as discussed in Ref. [39, 137, 155]. Its validity is diminished by the following factors: (a) shortness of separating distances at the stage of collisions; (b) field-induced torques that modify the interactions; (c) high flow velocities that modify $\hat{\mathbf{n}}$ [156], etc. Other types are possible, for example, Saturn rings with quadrupolar interactions[32, 157, 34, 158]. Different (dipolar and quadrupolar) meta-atoms can be combined in a quasi-equilibrium manner in the same LC matrix to yield complex structures [159, 160]. The term "quasi" is used to stress that since the energies involved in the assembly of colloids in LCs are typically two-three orders of magnitude larger than $k_BT$, the resulting structures are not truly equilibrium. By melting the LC and breaking the aggregates and then repeating the experiment again in the LC phase with dispersed particles, one would obtain new aggregates, in which only the overall geometry is preserved, but not the concrete positions of the "meta-atoms."

Anisotropic arrangements can be also staged at the interfaces, involving an isotropic medium and various LC phases, such as nematic [63-65, 161], cholesteric [58, 66], and smectic [59, 61, 162]. Trapping and dragging colloids by moving isotropic-LC interface offers another interesting mechanism of transport [163]. The moving isotropic-LC interface can be controlled thermally or by photo-induced trans-cis isomerization[164].

The anisotropy of particle interactions can be caused not only by the particle shape (as in an isotropic fluid [154]) and not only by the anisotropic host LC medium, as discussed above, but also by combining these two mechanisms. Velev's group demonstrated reconfigurable anisotropic clusters formed by metallo-dielectric Janus particles in an isotropic fluid, subject to the external electric field that polarizes the particles and creates non-centrosymmetric interactions among them[14], and aggregates formed by magnetic Janus particles under the action of magnetic and electric fields [165]. Some of these approaches can be staged in a LC; note that the dielectric anisotropy of the LC can be designed to adopt values in a broad range, including the zero value that would allow one to polarize the particles without realigning the director.

## 7. THERMAL EXPANSION

Materials expand when heated and contract when cooled because of temperature-induced changes in distances between molecules and atoms. In LCs, because of the coupling between the material flow and the director orientation, thermal expansion changes the structure by realigning the constituent molecules[166]. Figure 13(a) shows the effect for a capillary whose length and width are much larger that the thickness $h$ (measured along the $z$-axis). The nematic inside the capillary is aligned uniformly along the $z$-axis, $\hat{\mathbf{n}}_0 = (0,0,1)$. The two ends are open. The temperature field is uniform everywhere, but at some moment, it starts to increase (or decrease) steadily and uniformly, with a rate $\tau'$. The thermal expansion leads to displacement of the nematic with horizontal velocity $v_x \approx 6\beta\tau' x \frac{z}{h}\left(1 - \frac{z}{h}\right)$ that increases linearly with distance from the center of capillary; here $\beta$ is the coefficient of thermal expansion[166]. The flow along the $x$ axis realigns $\hat{\mathbf{n}}$ towards the $x$ axis. The viscous realigning torque is opposed by the elastic torque that tends to keep $\hat{\mathbf{n}}$ vertical. The balance of the two torques determines the non-uniform flow-induced director profile, Fig.13, expressed as the angle $\theta$ between the local $\hat{\mathbf{n}}$ and the $z$-axis: $\theta(x,z) \approx \beta\tau' xz \frac{|\alpha_2|}{K_3}\left(1 - \frac{z}{h}\right)\left(1 - 2\frac{z}{h}\right)$, where $\alpha_2$ is the Leslie viscosity coefficient. In order of magnitude, even modest temperature rates, $\tau' \sim 0.5^0\text{C/s}$, cause flow velocities of the order of 10 $\mu$m/s and director tilts by tens of degrees. Mirror symmetry of the director tilts with respect to the mid-plane of the cell, Fig.13(a), makes the uniaxial nematic appear similar to a biaxial nematic. In particular, conoscopic observations reveal a pattern of split isogyres, Fig.13(b). The splitting, however, is not a signature of a biaxial nematic in this case, but merely a result of the flow-induced tilt of the optic axis[166, 167]. The expansion-induced flows can carry particles and topological defects, Fig.13(b,c).

Nematic cells activated by varying temperature can be used for simultaneous thermo-mechanical and thermo-optical effects, such as transport of particles levitating in the nematic bulk with concomitant reorientation of optic axis around them. Simplicity of the phenomenon that does not require pumps nor even electrodes to produce dramatic optical and mechanical changes suggests that it might find applications in sensors, photonics, lab-on-a-chip, micro- and optofluidics. All these fields started to explore benefits offered by LC as a functional microfluidic and optofluidic medium [168-173].

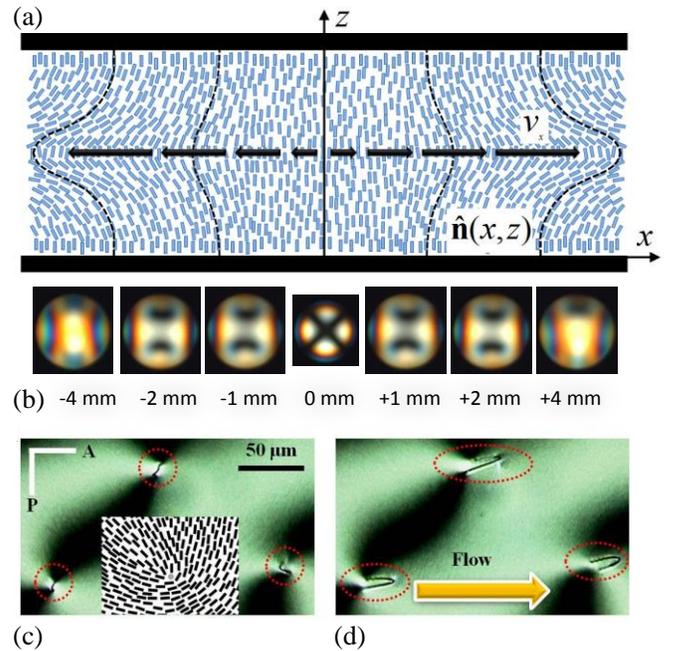

**Fig. 13.** *(a) Cross-section of a rectangular capillary with open ends filled with thermally expanding nematic. The horizontal velocity and director tilt increase linearly with the horizontal distance from the centre; (b) conoscopic patterns observed at different distances from the centre of a capillary with a thermally expanding uniaxial nematic LC[166] (c,d) thermal expansion causes displacement of disclinations [166].*

## 8. LIQUID CRYSTAL ENABLED ELECTROPHORESIS

Historically, the most popular technique of electrically-controlled transport of particles in an isotropic fluid is electrophoresis [174-177]. Under a uniform DC electric field, the particle moves with the velocity that depends linearly on the applied electric field; under normal conditions, a symmetric AC field with a zero time average produces no net displacement.

There is a growing interest in finding mechanisms that would allow one to manipulate particles with an AC driving, as with the latter, it is much easier to produce steady flows and to avoid undesirable electrochemical reactions. The nonlinear electrokinetic response was described for particles in an isotropic fluid, first by Murtsovkin et al [178-181], and then analysed systematically by Bazant and Squires [182-186], who predicted that asymmetric particles can show a quadratic dependence of the electrophoretic velocity on the applied electric field, in the so-called induced charge electrophoresis (ICEP). Experimentally, ICEP has been demonstrated for Janus metallo-dielectric particles by Velev et al [187].

Electrophoresis in LCs is studied much less. Replacement of an isotropic fluid with a LC should first of all result in anisotropy of the electrophoretic velocity with respect to $\hat{\mathbf{n}}_0$ because of the different Stokes drags [77, 79, 80, 138, 158, 188-192]. There are, however, some qualitative differences. Dierking et al [193] has reported on the electromigration of microspheres in nematic LCs and noticed that the particles move under the AC field in the direction perpendicular to the field and parallel to $\hat{\mathbf{n}}_0$; the velocity was linearly dependent on $E$. Ryzhkova, Podgornov and Haase [194] found a nonlinear (cubic) term in the dependence of $v$ on $E$, in addition to the classic linear term, i.e.,

$$v \propto \mu_1 E + \mu_3 E^3, \quad (6)$$

where $\mu_1$ and $\mu_3$ are the linear and third-order mobilities. Sikharulidze [195] has proposed to use electrophoresis in LCs for electrophoretic displays, as reviewed by Klein [196].

A different effect of LC-enabled electrophoresis (LCEEP) with the velocity or some of its components depending on the square of the field,

$$v \propto E^2,$$

has been described in Ref. [45, 67]. The relationship is similar to that in ICEP, but is applicable to absolutely symmetric spheres; the required symmetry breaking is provided by the LC medium.

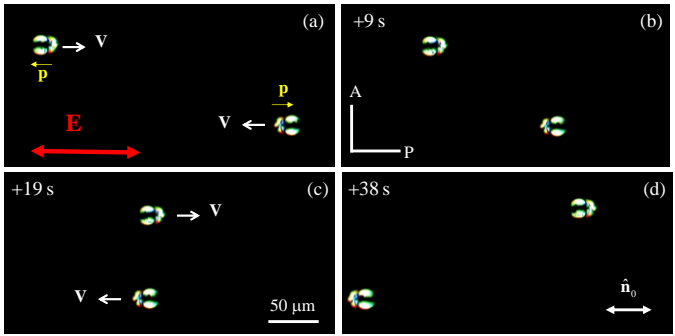

**Fig.14**. *Liquid crystal-enabled electrophoresis of two silica spheres with normal surface anchoring and opposite orientation of the elastic dipole $\mathbf{p}$ (directed from the hyperbolic hedgehog towards the sphere), driven by an AC electric field (30 $\mathrm{mV/\mu m}$) of frequency 1 Hz, in a nematic LC with vanishingly small dielectric anisotropy $\varepsilon_a = 0.03$.[45] Electrophoretic velocity $\mathbf{v}$ is antiparallel to the elastic dipole $\mathbf{p}$ of the director distortions around the sphere; $\mathbf{v}$ is collinear with field $\mathbf{E}$ and the overall director $\hat{\mathbf{n}}_0$. Polarized light microscopy with crossed polarizer and analyser.*

The difference between the quadratic electrophoretic response $v \propto E^2$ and any other form in which the velocity is an odd function of the electric field, Eq.(6), is principal. First, the dependence $v \propto E^2$ allows one to move particles even by a symmetric (for example, a sinusoidal) AC field with a zero time average, Fig.14. Second, since the blocking effect of free ions is diminished, LCEEP can produce steady flows that do not decay with time. Third, the dependence $v \propto E^2$ implies that the mechanism is not related to the particle's charge; LCEEP can thus carry particles of zero charge. Fourth, the relationship between the vectors $\mathbf{v}$ and $\mathbf{E}$ in an anisotropic medium should be generally of tensor nature, so that the velocity is not necessarily parallel to the applied field and can be, say, perpendicular to it.

The essence of LCEEP and its difference from the linear electrophoresis of charged particles can be illustrated by an experimental situation in Fig.15 that is similar to the one discussed in Section 3.3. Namely, one deals with the nematic of negative dielectric anisotropy $\varepsilon_a < 0$, in a planar cell $\hat{\mathbf{n}}_0 = (1,0,0)$ with the electric field directed perpendicular to $\hat{\mathbf{n}}_0 = (1,0,0)$. The difference is that the electric field is applied in the plane of the cell, Fig.15, $\mathbf{E} = (0, E_y, 0)$, rather than across the cell, as in Fig.6. In this geometry, the particles are moving in the plane of the cell, which allows one to trace their trajectories in details under the microscope. Since $\mathbf{E} \perp \hat{\mathbf{n}}_0$, the electric field does not reorient the director far away from the particle [67], Fig.15.

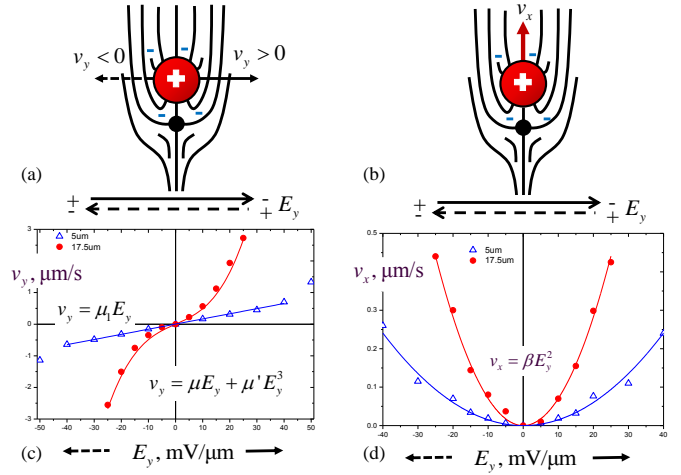

**Fig.15.** *(a) Scheme of a regular electrophoretic motion of a positively charged sphere in a nematic with $\varepsilon_a < 0$; the velocity is an odd function of the applied field; (b) nonlinear electrophoresis with the velocity growing as a square of the field; (c) linear dependence $v_y \propto E_y$ of the electrophoretic velocity in the direction of applied field for positively charged glass spheres of diameter 5 $\mu\mathrm{m}$; for larger spheres (diameter 17.5 $\mu\mathrm{m}$) there is also a cubic term in the dependency $v_y(E_y)$; (d) nonlinear LC-enabled electrophoresis in a direction perpendicular to the electric field, with velocity $v_x \propto E_y^2$; see text and Ref. [67] for details.*

The particles are small (microns) dielectric (glass) or metallic (gold) spheres. The spheres' surface anchoring is perpendicular and the director adopts a dipolar configuration, Figs.1(c), 2(a,b). If there is no voltage, the spheres in the LC experience Brownian motion with two independent self-diffusion coefficients $D_\parallel \propto \eta_\parallel$ and $D_\perp \propto \eta_\perp$, associated with the motion parallel and perpendicular to $\hat{\mathbf{n}}_0$, with two different effective viscosities, $\eta_\parallel$ and $\eta_\perp$, respectively, Fig.3.

Once the field is applied in the $xy$ plane of the cell, $\mathbf{E}=(0,E_y,0)$, perpendicularly to $\hat{\mathbf{n}}_0$, the glass spheres start to move. Charged glass spheres show two components of electrophoretic velocity in the $xy$ plane, Fig.15. The component $v_y$ parallel to the field grows linearly with the field if the spheres are relatively small (diameter 5 μm). For large spheres, this component acquires also a cubic term in its dependence of the applied electric field $E_y$, Fig.15 (a,c). The nonlinear behaviour of $v_y(E_y)$ is well described by an odd function with a cubic term in Eq.(6). Note that if the nematic is melted into an isotropic fluid, the qualitative behaviour of $v_y(E_y)$ remains intact. However, when the sphere is not charged, the velocity $v_y$ is zero; experimentally, this regime is observed when the glass spheres are replaced with gold spheres.

There is also a second velocity component $v_x$, perpendicular to $\mathbf{E}=(0,E_y,0)$, that is not zero for both charged and un-charged spheres moving in the LC, Fig.15(b,d). This velocity vanishes when the LC is melted into an isotropic fluid. The dependence $v_x(E_y)$ is quadratic:

$$v_x = \beta E_y^2 , \quad (7)$$

where $\beta$ is the nonlinear coefficient, the sign of which changes with the reversal of the elastic dipole $\mathbf{p}$. The LCEEP is observed not only in the LC with $\varepsilon_a<0$, Fig.15(b,d), but also in nematics with $\varepsilon_a>0$; in this case, the elastic dipole $\mathbf{p}$ and the direction of motion are parallel or antiparallel to the field direction $\mathbf{E}\parallel\hat{\mathbf{n}}_0$, Fig.15. The dependence (7) implies that the particles in LCEEP can be driven by the AC field, Figs.14 and 15.

When the LC-enabled electrophoresis is staged in a LC with $\varepsilon_a\le 0$ and driven by the field $\mathbf{E}=(0,0,E_z)$ that is perpendicular to the plane of the cell, then the $\mathbf{E}\perp\hat{\mathbf{n}}_0$ is fulfilled for any director $\hat{\mathbf{n}}_0(x,y)$ in the plane of the cell. This degeneracy can be used to design 2D director "rails" $\hat{\mathbf{n}}_0(x,y)$ to guide the particles transport along curvilinear paths[45]. In Fig.16, the director is shaped as concentric circles set by circular buffing of the alignment layers. Under an AC driving field, the spheres move circularly, either clockwise or counter-clockwise, depending on the direction of $\mathbf{p}$. Since the field polarity is alternating, there is no net shift along the $z$-axis.

Using both AC and DC components of the vertical field $\mathbf{E}=(0,0,E_z)$ acting on a nematic LC with $\varepsilon_a\le 0$, one can design practically any 3D trajectory of the electrophoretic particle. The $z$-location can be controlled by the linear/cubic electrophoresis driven by a DC field, Fig.6, while the in-plane location of the particle can be controlled by an AC field and an appropriately designed director.

The quadratic AC and DC electrophoresis is observed for spherical particles only when the director distortions around them are of polar symmetry, Fig.1(c), 2. If the hedgehog is spread into an equatorial disclination loop, the symmetry is quadrupolar, Fig.1(d), and the quadratic electrophoresis vanishes[45]. Thus the LCEEP is rooted in the type of director distortions which violate the symmetry of the particle through the associated director "coat".

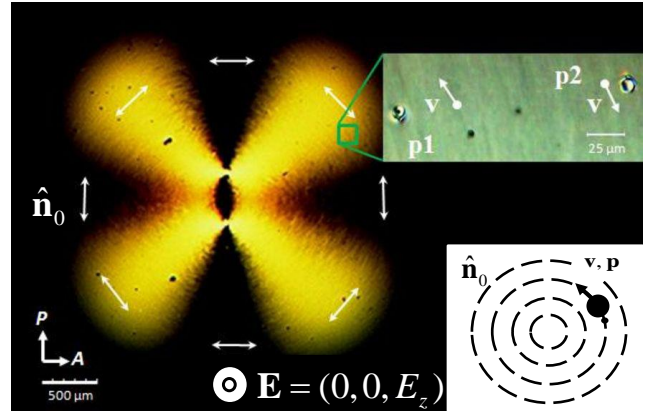

**Fig.16.** *Circular electrophoretic motion[45] of glass spheres in the nematic with $\varepsilon_a<0$. The director alignment is circular. The electric field is normal to the plane of view. The particles move clockwise and counter-clockwise, depending on the direction of elastic dipole, in the plane perpendicular to the electric field.*

In the experiments with a nonzero dielectric anisotropy, one might wonder whether the electric-field induced modifications of the director in the vicinity of the particle can contribute to the propulsion of the spheres. They very well might, but the effect is not decisive, as in mixtures with $\varepsilon_a=0$ one still observes an LCEEP with the velocity that is of the same order as the velocity in a LC with nonzero $\varepsilon_a$.

To understand why the LCEEP does not depend on the polarity of $\mathbf{E}$, consider an uncharged sphere in a nematic with $\varepsilon_a=0$. The nematic contains an equal amount of positive and negative charges. Once a uniform electric field is applied, these mobile ions start to move in opposite directions, driven by the Coulomb force. Ionic mobility $\mu_\parallel$ parallel to the local director $\hat{\mathbf{n}}$ is different from the mobility $\mu_\perp$ perpendicular to it. Also different are the effective viscosities $\eta_\parallel$ and $\eta_\perp$ for local LC fluid motion parallel and perpendicular to the director $\hat{\mathbf{n}}$. The associated electro-osmotic flows near the top and bottom parts of the sphere in Fig.17(a) are thus generally different because the pattern of $\hat{\mathbf{n}}$ in a dipole-like structure of a colloid and an accompanying hedgehog is different. The mirror symmetry of electro-osmotic flows with respect to the plane shown by the dashed line in Fig.17 (a) is thus broken. If the flows are stronger in the bottom part, as in Fig.17(a), the sphere is expected to move upwards. Field reversal does not change the direction of LCEEP velocity, since it does not change the symmetry of director distortions, Fig.17(a,b): Once the hedgehog is formed near the north pole of the sphere, it cannot be moved to the south pole. In materials with a big difference of ionic mobility for positive and negative ions, more subtle effects can be added to the consideration, but they do not change the qualitative picture of LC-enabled electrophoresis. The broken symmetry argument suggests that the electrophoretic velocity grows as $v_x \propto E_y^2$ and maintains the direction dictated by the elastic dipole p, Fig.17(a,b). The relationship between the electgrophoretic velocity and p should depend on a number of factors that influence the concrete director configuration around the sphere and the pathway of moving ions and fluid elements through this configuration. Among these factors the most obvious are anisotropic mobility of ions and viscosity, but one should also consider the dielectric and flexoelectric anisotropy of the nematic, as they modify the director each time the electric field is switched on or off. Experimentally[45, 67], particles move parallel to p in LCs with a large $|\varepsilon_a|$, Fig.16, and antiparallel to p when $\varepsilon_a \approx 0$, Fig.14.

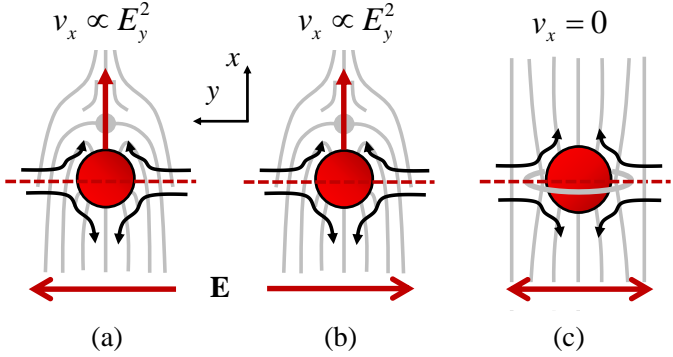

**Fig.17.** *(a,b) Scheme of LC-enabled electrophoresis of a sphere with dipolar director distortions in a nematic with $\varepsilon_a = 0$; the symmetry of electro-osmotic flows is broken with respect to the dashed line; both polarities of the electric field cause the same direction of electrophoresis, $v_x \propto E_y^2$; (c) quadrupolar symmetry of the director field does not result in electrophoretic motion of the sphere.*

If a colloidal sphere with normal surface anchoring features an equatorial Saturn ring director configuration instead of the dipolar director, then such a structure is of a quadrupolar symmetry, and the LCEEP mechanism of motion described above should be impossible, Fig.17(c), which agrees with the experiment [45]. To further advance the understanding of LCEEP mechanisms, it would be of interest to (1) map the fluid velocity pattern around the particles; the symmetry of these patterns would be directly related to the electrophoretic ability of the colloid; and (2) vary the ratio of conductivities $\mu_\parallel / \mu_\perp$ and effective viscosities $\eta_\parallel / \eta_\perp$ in order to alter the polarity of the electrophoretic transport. Using asymmetric particles in a LC-enabled electrophoresis would also be of interest as in this case $\mathfrak{p} \neq 0$ for any surface anchoring, including the tangential one; the asymmetric particles can be thus driven by both ICEEP and ICEP mechanisms.

LCEEP is similar to ICEP of asymmetric particles (such as metallo-dielectric Janus spheres) moving in an isotropic fluid, as in both cases the principal dependence of velocity on the applied field is quadratic, $v \propto E^2$. The difference is that in the first case, it is the symmetry of the medium that is broken, while in the second case, it is the symmetry of the particle that is broken. Another difference is that once created, the elastic dipole **p** of a particle in a LC cannot change its direction easily (without dramatic intrusions such as melting of the nematic). In the case of Janus particles in ICEP, the structural dipole (induced by polarization of the electric field) and the direction of motion are orthogonal to each other[187]. As a result, the velocity vector can tumble (remaining perpendicular to the field) and the Janus spheres can change the polarity of their motion.

The experiments demonstrate the tensor character of the relationship between the velocity and the field in LCCE. The fact that the particles can move perpendicularly to **E** in nematics with $\varepsilon_a < 0$ means that the driving voltage can be applied across the thickness of the channel (few μm) rather than along the pathway (several cm) in electrophoretic devices; it allows one to use modest voltage sources to achieve high fields, which might be especially beneficial for portable devices. The possibility of moving the particle in different directions without altering the direction of the field is also remarkable. The field and frequency dependencies of LCEEP indicate that one can combine two differently oriented driving fields with different frequencies for a better control of the overall particle's trajectory. Further diversification can be achieved by using LCs with distorted director and by using LCs of zero dielectric anisotropy so that the electric field does not cause the director reorientation regardless of the mutual orientation of $\hat{\mathbf{n}}_0$ and **E**.

The study of LCEEP and other nonlinear electrokinetic effects in LCs is in its infancy, but the richness of phenomena already observed suggests that the field will attract more researchers. Potential directions of future research are exploration of lyotropic LCs as a carrier medium, transport of particles of non-spherical shape, motion of particles that are liquid or soft rather than solid, etc. Hernàndez-Navarro et al[197] recently demonstrated that LCEEP can be used to transport water microdroplets; these can contain chemicals or drugs and thus serve as controllable microreactors with micrometer-precise delivery.

## 9. QUINCKE ROTATION AND TRANSPORT

The electric field can cause not only translation of the inclusions in LCs but also their rotations, caused, for example, by the Quincke effect. Quincke rotation is defined as a spinning motion of a dielectric particle neutrally buoyant in an isotropic fluid powered by a DC electric field. The necessary condition of rotation is that the charge relaxation time of the particles is longer than that of the fluid medium. If this is the case, the field-induced polarization of the particle is anti-parallel to the applied electric field, Fig.18. Such an orientation is unstable and, if the field is sufficiently high to overcome the viscous friction, one observes a steady rotation with a constant angular velocity [198] $\Omega = \pm \frac{1}{\tau_{MW}} \sqrt{\frac{E^2}{E_c^2} - 1}$, where $\tau_{MW} = \varepsilon_0 (\varepsilon_p + 2\varepsilon_m)/(\sigma_p + 2\sigma_m)$ is the Maxwell-Wagner relaxation time, $\varepsilon_p (\varepsilon_m)$ and $\sigma_p (\sigma_m)$ are the relative dielectric permittivities and the conductivities of the particle (liquid medium), and $E_c$ is the critical field. In absence of shear flow, the spin direction of an isolated spherical particle is arbitrary in the plane perpendicular to the electric field.

Quincke rotation has been reported by Jákli et al [60, 199] for small spheres and cylinders placed in the nematic and smectic A LCs. Heating the samples and melting the LC does not stop the rotation, as expected in light of the general mechanism of rotation, but changes its angular velocity, which is also a natural result since the threshold field depends on the medium properties [198], such as the viscosity $\eta$:

$E_c = \eta \frac{\sigma_m}{\varepsilon_0^2 \varepsilon_m \varepsilon_p}$. In the case of a smectic A LC, the relevant viscosity is the Miesowicz viscosity $\eta_a$, because the shear plane is perpendicular to the director[18]. Therefore, the Quincke rotation can be used to determine $\eta_a$. For glass particles in the smectic A phase of 8CB, it was found that $E_c = 0.6 \text{ V/}\mu\text{m}$, $\Omega = 30 \text{ s}^{-1}$ at $E = 2.5 \text{ V/}\mu\text{m}$; the experimentally determined field dependence of the angular velocity resulted in $\eta_a = 1.6 \text{ Pa s}$.

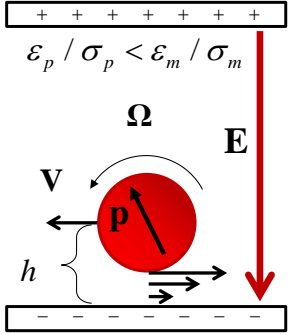

**Fig.18**. *Quincke rotation of a sphere powered by DC electric field; **p** is the electric polarization caused by the applied field **E**. If the sphere is not located in the mid-plane, hydrodynamic interaction with the nearest wall causes translational motion along the wall with the velocity **V** proportional to the angular velocity of the Quincke rotation.*

The ordered nature of LC media leads also to new qualitative features of the Quincke rotation[60]. At some threshold field $E_{tr} > E_c$, the Quincke rotators placed in a bounded cell of either a LC or isotropic fluid, start to move, mostly in the plane perpendicular to the electric field, Fig.19. The effect cannot be explained by backflow and the nonlinear electrophoresis effects described in previous sections, since the transport is observed also in the isotropic melt. It was proposed that the transport is caused by hydrodynamic interaction between the rotating particle and the bounding wall of the sample[60].

The hydrodynamic interaction is relevant when the particle-wall separation is much shorter than the viscous penetration length $\zeta = \sqrt{\frac{\eta}{\rho\Omega}}$, which for typical viscosity, density and angular velocity is large (~1 mm) as compared to the typical cell thickness. If the sphere spins at a distance $h < \zeta$ from the wall, the velocity gradient between the wall and the sphere is much steeper (and thus the viscous stress is larger) than in the rest of the space, Fig.18, so there is a force pushing the sphere along the wall, perpendicular to the axis of spinning. By balancing the torques and forces acting on the Quincke rotator near the wall, one finds the velocity of translation in the direction perpendicular to both the applied electric field and the spinning axis [60], $V = \frac{1}{8} R\Omega \left(\frac{R}{h}\right)^4$. For fields on the order of 10 V/μm, the velocities might be very high, (40-50) V/μm.

In smectic A samples with air bubbles, the particles are strongly trapped in the meniscus region, at the grain boundary separating the regions of differently tilted smectic layers. In the applied electric field, the particles quickly orbit the air bubble, Fig.19(a,b), remaining within the meniscus region because of the strong elastic trapping forces. In the nematic phase, the elastic forces caused by the director distortions are weaker than in the smectic case and the Quincke rotators are released, Fig.19(c). Transport of Quincke rotators represents an interesting example of how the linear stimulus (DC electric field) is first converted into spinning and then the spinning is converted into linear and orbital translations.

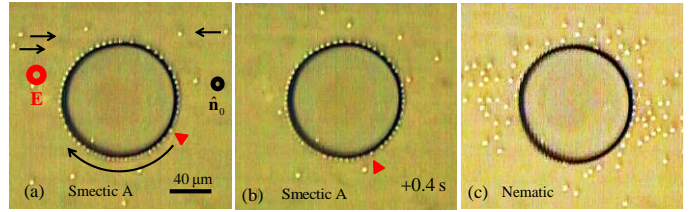

**Fig.19**. *Glass spheres of diameter 4.5 μm in a homeotropic smectic A (a,b) and nematic (c) slabs with an air bubble in the centre. The spheres experience Quincke rotation, unidirectional and orbital motion (indicated by black arrows in (a)). One of the particles is labelled by a triangle to highlight the orbital motion around the air bubble meniscus. In the smectc A, the spheres are strongly trapped at the meniscus, while in the nematic, the trapping force is much smaller. The electric field is perpendicular to the plane of figure. Microphotographs courtesy of A. Jákli.*

## 10. CONCLUSIONS

Exploration of the properties of LCs as a functional medium for transport and assembly of "meta-atoms" that would be capable to support reconfigurable materials and systems has started only recently. The field develops in parallel with a much larger effort focused on isotropic fluids as a medium in which micro- and nanometer particles show individual and collective dynamics and ability to form various self-assembled patterns, for example, in response to external electromagnetic fields, as reviewed recently by Aranson [6] and by Dobnikar et al. [200]. There are unifying features of dynamics in isotropic and anisotropic fluids, such as absence of inertia at the scale of microns and smaller or the possibility to polarize the particles by the external fields. The differences, evident even in statics, such as LC-enabled levitation, show an extraordinary potential of the anisotropic fluids in designing soft active metamatter.

LC-enabled elastic levitation mitigates the detrimental effect of gravity that plagues attempts to grow 3D structure in isotropic media through the bottom-up approaches, by offering a mechanism of elastic repulsion from the boundaries by which the particles can resist sedimentation. The dynamic phenomena show even a richer spectrum of opportunities, because of (1) long-range orientational order of the LCs that can be locally controlled or modified; (2) coupling of material flows to the director orientation; (3) anisotropic character of physical properties of the LCs, such as dielectric permittivity and mobility of ions.

LCs feature long-range orientational order. This order can be modified by various means, from temperature to electromagnetic fields, from boundary conditions to introduction of foreign molecules or particles. By creating the spatial gradients of the order parameter, either in terms of the director field, or in terms of the scalar order parameter, or both, one sets up the dispersed particles into directed motion towards a location that corresponds to an equilibrium or a metastable state dictated by the balance of anisotropic elastic and surface interactions. Examples considered in the review included director gradients near the cores of topological defects, director gradients caused by the electric field in the dielectrically realigning LC and locally induced gradients in the degree of molecular order. Yet another quintessentially LC mechanism of transport can be based on dielecrophoresis caused by spatial variations of permittivity of the distorted LC.

Coupling of the material flow and director reorientation in LCs is typically associated with undesirable effects such as backflow in LC displays. In the context of colloidal transport, this coupling can be useful, as the periodic director reorientation sets the LC and dispersed particles into steady motion; the cells can be tuned to produce various morphologies of colloidal aggregates created through inelastic scattering of particles. Another example of the flow-director coupling is the reorientation of the director as a result of thermal expansion that can also be used in colloidal transport.

Anisotropy of LCs sets up new facets of electrophoresis. The LCEEP is characterized by the quadratic dependence of the velocity on the applied electric field. The electrophoretic velocity can be directed not only parallel to the field but also perpendicularly to it, because of the tensor character of LCEEP. LCEEP allows one to use AC driving, to create steady flows, to move perfectly symmetric spheres and particles that are deprived of any surface charges, to design the particles trajectories by creating a spatially (or temporarily) varying director field. A similar opportunity of designing trajectories of colloidal transport is offered by translations of Quincke rotators that can be trapped in certain regions of the samples and moved around, as illustrated in Fig.19. The review is limited by consideration of mostly individual particles; phenomena associated with interactions of moving particles and collective dynamic behaviour remain practically unexplored. Another limitations is that the review focuses on thermotropic LCs and does not discuss much their lyotropic counterparts[201-203]. Some of the presented features and mechanisms should be relevant to both types of LCs. For example, lyotropic LCs also show a preferred surface alignment at the interfaces. Typically, in lyotropics formed by micelles of surfactant molecules, surface orientation is determined by the excluded volume consideration [204] (thus the nematic formed by worm-like micelles prefers to align parallel to the substrate[84]). Lyotropic chromonic LCs can be aligned in both tangential and homeotropic fashion, depending on the nature of the substrate [27]. The surface anchoring is sufficiently strong to cause director distortions around colloidal inclusions, as evidenced by Fig.20 in which spheres in a chromonic LC show either strong tangential or strong perpendicular anchoring and create topological defects such as two surface boojums, Fig.20(a) or a hyberbolic hedgehog, Fig.20(b). Thus some aspects of static (anisotropic interactions, levitation), and dynamic (anisotropic Brownian motion [84]) behaviour of colloids in lyotropics should be similar to those in thermotropics. However, electric field induced effects are expected to be different, first of all because most lyotropic LCs contain a large number of ions. Ionic motion in the electric field drags the LC; dielectric reorientation of the director is practically non-existent. Furthermore, high concentration of ions means that the electric double layers around colloids in lyotropic LCs are very thin; the latter might hinder effects such as nonlinear electrophoresis. Generally, the interest to the colloidal interactions and dynamics in thermotropic LCs is expected to expand to the lyotropic domain.

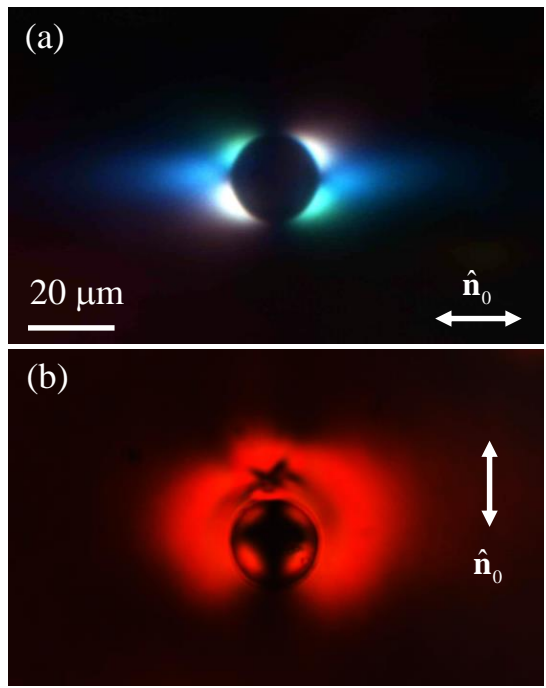

**Fig.20.** *Polarizing microscope texture of a silica sphere in a nematic chromonic liquid crystal (a) disodium chromoglycate dispersion in water, 14 wt%; tangential surface anchoring, bipolar structure with two boojums; (b) Sunset Yellow dispersed in water (33wt%) at 40°C. The sphere is treated with N,N-didecyl-N-methyl-(3-trimethoxysilylpropyl) ammonium chloride to orient the director perpendicularly to its surface. Sample preparation and microphotography by Israel Lazo.*

The richness of colloidal transport phenomena in LCs promises to develop into a rich toolbox that would facilitate design and manufacture of soft active metamatter. However, much more needs to be learned about the underlying mechanisms before applications might become a reality.

## Acknowledgements

The review is based on the research performed by former and current graduate students Y-K. Kim, I. Lazo, O. Pishnyak, I. Senyuk, I. Smalyukh, and D. Voloshchenko and in collaboration with A. Jákli, J. Kelly, S. Shiyanovskii, and S.-P. Tang. I am thankful to all the participants of the SAMM-2012 and -2013 projects for illuminating discussions that helped to select the material for the review. I gratefully acknowledge the hospitality of the Isaac Newton Institute, Cambridge, UK, where part of this review was written. This research was supported by DOE grant DE-FG02-06ER 46331 in the studies of electric-field induced backflow and dynamics of cluster formation and NSF DMR 1104850 in studies of electrophoresis.

## References

1. P. G. De Gennes (1992) *Science* **256**, 495-497.
2. F. Ilievski, A. D. Mazzeo, R. E. Shepherd, X. Chen, and G. M. Whitesides (2011) *Angewandte Chemie-International Edition* **50**, 1890-1895.
3. S. Ramaswamy (2010) *Annual Review of Condensed Matter Physics, Vol 1* **1**, 323-345.
4. F. C. Mackintosh and M. E. Cates (2011) *Soft Matter* **7**, 3050-3051.
5. M. C. Marchetti, J.-F. Joanny, S. Ramaswamy, T. B. Liverpool, J. Prost, Madan Rao, and R. Aditi Simha (2013) *Reviews of Modern Physics* **85**, 1143-1189.
6. I. S. Aranson (2013) *Physics-Uspekhi* **56**, 79-92.
7. W. S. Cai, U. K. Chettiar, A. V. Kildishev, and V. M. Shalaev (2007) *Nat Photonics* **1**, 224-227.
8. G. M. Whitesides and B. Grzybowski (2002) *Science* **295**, 2418-2421.
9. A. Sokolov, M. M. Apodaca, B. A. Grzybowski, and I. S. Aranson (2010) *P Natl Acad Sci USA* **107**, 969-974.
10. A. Würger (2010) *Rep Prog Phys* **73**, 126601.
11. Y. T. Ma, A. Bhattacharya, O. Kuksenok, D. Perchak, and A. C. Balazs (2012) *Langmuir* **28**, 11410-11421.
12. I. B. Burgess, L. Mishchenko, B. D. Hatton, M. Kolle, M. Loncar, and J. Aizenberg (2011) *J Am Chem Soc* **133**, 12430-12432.
13. J. T. B. Overvelde, S. Shan, and K. Bertoldi (2012) *Adv Mater* **24**, 2337-2342.
14. S. Gangwal, A. Pawar, I. Kretzschmar, and O. D. Velev (2010) *Soft Matter* **6**, 1413-1418.
15. A. Snezhko, M. Belkin, I. S. Aranson, and W. K. Kwok (2009) *Phys Rev Lett* **102**, 118103.
16. M. K. Gupta, D. D. Kulkarni, R. Geryak, S. Naik, and V. V. Tsukruk (2013) *Nano Lett* **13**, 36-42.
17. P. G. De Gennes and J. Prost The Physics of Liquid Crystals (Clarendon Press, Oxford 1993) p 598.
18. M. Kleman and O. D. Lavrentovich Soft Matter Physics: An Introduction (Springer, New York 2003) p 638.
19. L. M. Blinov and V. G. Chigrinov Electrooptic Effects in Liquid Crystal materials (Springer, New York 1994) p 540.
20. B. Jerome (1991) *Rep Prog Phys* **54**, 391-451.
21. D. S. Miller, R. J. Carlton, P. C. Mushenheim, and N. L. Abbott (2013) *Langmuir* **29**, 3154-3169.
22. Y. A. Nastishin, R. D. Polak, S. V. Shiyanovskii, and O. D. Lavrentovich (1999) *Appl Phys Lett* **75**, 202-204.
23. M. Vilfan, A. Mertelj, and M. Čopič (2002) *Physical Review E* **65**, 041712.
24. I. I. Smalyukh and O. D. Lavrentovich (2003) *Phys Rev Lett* **90**, 085503.
25. M. Ruths and B. Zappone (2012) *Langmuir* **28**, 8371-8383.
26. V. G. Nazarenko and O. D. Lavrentovich (1994) *Physical Review E* **49**, R990-R993.
27. V. G. Nazarenko, O. P. Boiko, H. S. Park, O. M. Brodyn, M. M. Omelchenko, L. Tortora, Y. A. Nastishin, and O. D. Lavrentovich (2010) *Phys Rev Lett* **105**, 017801.
28. F. Brochard and P. G. D. Gennes (1970) *J Phys-Paris* **31**, 691-&.
29. F. H. Li, O. Buchnev, C. Il Cheon, A. Glushchenko, V. Reshetnyak, Y. Reznikov, T. J. Sluckin, and J. L. West (2006) *Phys Rev Lett* **97**.
30. F. Li, O. Buchnev, C. Il Cheon, A. Glushchenko, V. Reshetnyak, Y. Reznikov, T. J. Sluckin, and J. L. West (2007) *Phys Rev Lett* **99**, 219901.
31. L. M. Lopatina and J. V. Selinger (2009) *Phys Rev Lett* **102**, 197802.
32. O. V. Kuksenok, R. W. Ruhwandl, S. V. Shiyanovskii, and E. M. Terentjev (1996) *Physical Review E* **54**, 5198-5203.
33. P. Poulin, H. Stark, T. C. Lubensky, and D. A. Weitz (1997) *Science* **275**, 1770-1773.
34. Y. D. Gu and N. L. Abbott (2000) *Phys Rev Lett* **85**, 4719-4722.
35. M. Škarabot, M. Ravnik, S. Zumer, U. Tkalec, I. Poberaj, D. Babic, and I. Musevic (2008) *Physical Review E* **77**, 061706.
36. H. Stark (2002) *Physical Review E* **66**, 032701.
37. H. Stark (1999) *European Physical Journal B* **10**, 311-321.
38. S. V. Shiyanovskii, T. Schneider, I. I. Smalyukh, T. Ishikawa, G. D. Niehaus, K. J. Doane, C. J. Woolverton, and O. D. Lavrentovich (2005) *Physical Review E* **71**, 020702.
39. V. M. Pergamenshchik and V. A. Uzunova (2010) *Condensed Matter Physics* **13**, 33602.
40. V. M. Pergamenshchik and V. A. Uzunova (2011) *Physical Review E* **83**, 021701.
41. C. P. Lapointe, T. G. Mason, and I. I. Smalyukh (2009) *Science* **326**, 1083-1086.
42. B. Senyuk, Q. K. Liu, S. L. He, R. D. Kamien, R. B. Kusner, T. C. Lubensky, and I. I. Smalyukh (2013) *Nature* **493**, 200-205.
43. C. P. Lapointe, K. Mayoral, and T. G. Mason (2013) *Soft Matter* **9**, 7843-7854.
44. O. P. Pishnyak, S. Tang, J. R. Kelly, S. V. Shiyanovskii, and O. D. Lavrentovich (2007) *Phys Rev Lett* **99**, 127802.
45. O. D. Lavrentovich, I. Lazo, and O. P. Pishnyak (2010) *Nature* **467**, 947-950.
46. M. Yada, J. Yamamoto, and H. Yokoyama (2004) *Phys Rev Lett* **92**, 185501.
47. I. I. Smalyukh, A. N. Kuzmin, A. V. Kachynski, P. N. Prasad, and O. D. Lavrentovich (2005) *Appl Phys Lett* **86**, 021913.
48. Smalyukh, Ii, O. D. Lavrentovich, A. N. Kuzmin, A. V. Kachynski, and P. N. Prasad (2005) *Phys Rev Lett* **95**, 157801.
49. I. Muševič, M. Škarabot, U. Tkalec, M. Ravnik, and S. Žumer (2006) *Science* **313**, 954-958.
50. M. R. Mozaffari, M. Babadi, J. Fukuda, and M. R. Ejtehadi (2011) *Soft Matter* **7**, 1107-1113.
51. M. Škarabot, M. Ravnik, S. Zumer, U. Tkalec, I. Poberaj, D. Babic, N. Osterman, and I. Muševič (2008) *Physical Review E* **77**, 031705.
52. V. S. R. Jampani, M. Škarabot, S. Čopar, S. Žumer, and I. Muševič (2013) *Phys Rev Lett* **110**, 177801.
53. M. A. Gharbi, D. Seč, T. Lopez-Leon, M. Nobili, M. Ravnik, S. Žumer, and C. Blanc (2013) *Soft Matter* **9**, accepted.
54. I. Muševič and U. Tkalec (2013) *Soft Matter* **9**, accepted.
55. T. Araki, F. Serra, and H. Tanaka (2013) *Soft Matter* **9**, accepted.

56. C. Blanc, D. Coursault, and E. Lacaze (2013) *Liquid Crystals Reviews* **1**, to be published.
57. T. C. Lubensky, D. Pettey, N. Currier, and H. Stark (1998) *Phys Rev E* **57**, 610-625.
58. D. Voloschenko, O. P. Pishnyak, S. V. Shiyanovskii, and O. D. Lavrentovich (2002) *Physical Review E* **65**, 060701.
59. D. K. Yoon, M. C. Choi, Y. H. Kim, M. W. Kim, O. D. Lavrentovich, and H. T. Jung (2007) *Nat Mater* **6**, 866-870.
60. A. Jákli, B. Senyuk, G. X. Liao, and O. D. Lavrentovich (2008) *Soft Matter* **4**, 2471-2474.
61. Y. H. Kim, D. K. Yoon, H. S. Jeong, O. D. Lavrentovich, and H. T. Jung (2011) *Adv Funct Mater* **21**, 610-627.
62. K. Higashiguchi, K. Yasui, M. Ozawa, K. Odoi, and H. Kikuchi (2012) *Polymer Journal* **44**, 632-638.
63. I. I. Smalyukh, S. Chernyshuk, B. I. Lev, A. B. Nych, U. Ognysta, V. G. Nazarenko, and O. D. Lavrentovich (2004) *Phys Rev Lett* **93**, 117801.
64. A. B. Nych, U. M. Ognysta, V. M. Pergamenshchik, B. I. Lev, V. G. Nazarenko, I. Muševič, M. Škarabot, and O. D. Lavrentovich (2007) *Phys Rev Lett* **98**, 057801.
65. M. Tasinkevych and D. Andrienko (2010) *Condensed Matter Physics* **13**, 33603.
66. J. S. Lintuvuori, A. C. Pawsey, K. Stratford, M. E. Cates, P. S. Clegg, and D. Marenduzzo (2013) *Phys Rev Lett* **110**, 187801.
67. I. Lazo and O. D. Lavrentovich (2013) *Phil. Trans. R. Soc. A* **371**, 20120255.
68. O. P. Pishnyak, S. Tang, J. R. Kelly, S. V. Shiyanovskii, and O. D. Lavrentovich (2009) *Ukr. J. Phys.* **54**, 101-108.
69. V. M. Pergamenshchik and V. A. Uzunova (2009) *Physical Review E* **79**, 021704.
70. T. G. Sokolovska and G. N. Patey (2009) *J Phys-Condens Mat* **21**, 245105.
71. S. B. Chernyshuk, O. M. Tovkach, and B. I. Lev (2012) *Physical Review E* **85**, 011706.
72. C. Lapointe, A. Hultgren, D. M. Silevitch, E. J. Felton, D. H. Reich, and R. L. Leheny (2004) *Science* **303**, 652-655.
73. C. P. Lapointe, D. H. Reich, and R. L. Leheny (2008) *Langmuir* **24**, 11175-11181.
74. T. A. Wood, J. S. Lintuvuori, A. B. Schofield, D. Marenduzzo, and W. C. K. Poon (2011) *Science* **334**, 79-83.
75. W. T. Coffey, Y. P. Kalmykov, and J. T. Waldron The Langevin Equation: With Applications in Physics, Chemistry and Electrical Engineering (World Scientific, Singapore 1996) p 413.
76. A. Einstein (1905) *Ann. Phys. (Leipzig)* **17**, 549.
77. R. W. Ruhwandl and E. M. Terentjev (1996) *Physical Review E* **54**, 5204-5210.
78. H. Stark and D. Ventzki (2002) *Europhys Lett* **57**, 60-66.
79. H. Stark, D. Ventzki, and M. Reichert (2003) *J Phys-Condens Mat* **15**, S191-S196.
80. J. C. Loudet, P. Hanusse, and P. Poulin (2004) *Science* **306**, 1525-1525.
81. G. M. Koenig, R. Ong, A. D. Cortes, J. A. Moreno-Razo, J. J. De Pablo, and N. L. Abbott (2009) *Nano Lett* **9**, 2794-2801.
82. M. Skarabot and I. Musevic (2010) *Soft Matter* **6**, 5476-5481.
83. J. A. Moreno-Razo, E. J. Sambriski, G. M. Koenig, E. Diaz-Herrera, N. L. Abbott, and J. J. De Pablo (2011) *Soft Matter* **7**, 6828-6835.
84. F. Mondiot, J. C. Loudet, O. Mondain-Monval, P. Snabre, A. Vilquin, and A. Wurger (2012) *Physical Review E* **86**, 010401.
85. D. Abras, G. Pranami, and N. L. Abbott (2012) *Soft Matter* **8**, 2026-2035.
86. J. Sprakel, J. Van Der Gucht, M. a. C. Stuart, and N. a. M. Besseling (2008) *Phys Rev E* **77**, 061502.
87. I. Y. Wong, M. L. Gardel, D. R. Reichman, E. R. Weeks, M. T. Valentine, A. R. Bausch, and D. A. Weitz (2004) *Phys Rev Lett* **92**, 178101.
88. M. M. Alam and R. Mezzenga (2011) *Langmuir* **27**, 6171-6178.
89. X. L. Wu and A. Libchaber (2000) *Phys Rev Lett* **84**, 3017-3020.
90. X. L. Wu and A. Libchaber (2001) *Phys Rev Lett* **86**, 557-557.
91. A. Ott, J. P. Bouchaud, D. Langevin, and W. Urbach (1990) *Phys Rev Lett* **65**, 2201-2204.
92. Y. Gambin, G. Massiera, L. Ramos, C. Ligoure, and W. Urbach (2005) *Phys Rev Lett* **94**.
93. R. Ganapathy, A. K. Sood, and S. Ramaswamy (2007) *Epl-Europhys Lett* **77**.
94. R. Angelico, A. Ceglie, U. Olsson, G. Palazzo, and L. Ambrosone (2006) *Phys Rev E* **74**.
95. M. Pumpa and F. Cichos (2012) *J Phys Chem B* **116**, 14487-14493.
96. H. N. W. Lekkerkerker and R. Tuinier Colloids and the Depletion Interactions (Springer, Dordrecht 2011).
97. M. E. Cates and M. R. Evans eds (2000) *Soft and fragile matter: Nonequilibrium dynamics, metastability and flow* (Edinburgh University, Edinburgh), p 394.
98. S. B. Chernyshuk and B. I. Lev (2011) *Physical Review E* **84**, 011707.
99. I. I. Smalyukh, S. V. Shiyanovskii, and O. D. Lavrentovich (2001) *Chemical Physics Letters* **336**, 88-96.
100. Y. D. Yin, Y. Lu, B. Gates, and Y. N. Xia (2001) *J Am Chem Soc* **123**, 8718-8729.
101. Y. A. Vlasov, X. Z. Bo, J. C. Sturm, and D. J. Norris (2001) *Nature* **414**, 289-293.
102. K. A. Mirica, F. Ilievski, A. K. Ellerbee, S. S. Shevkoplyas, and G. M. Whitesides (2011) *Adv Mater* **23**, 4134-4140.
103. W. B. Russel, D. A. Saville, and W. R. Schowalter Colloidal Dispersions (Cambridge University Press, Cambridge 1989) p 526.
104. D. Pires, J. B. Fleury, and Y. Galerne (2007) *Phys Rev Lett* **98**, 247801.
105. J. C. Loudet, P. Barois, and P. Poulin (2000) *Nature* **407**, 611-613.
106. D. Coursault, J. Grand, B. Zappone, H. Ayeb, G. Levi, N. Felidj, and E. Lacaze (2012) *Adv Mater* **24**, 1461-1465.
107. T. Araki, M. Buscaglia, T. Bellini, and H. Tanaka (2011) *Nat Mater* **10**, 303-309.
108. S. Čopar, N. A. Clark, M. Ravnik, and S. Žumer (2013) *Soft Matter* **9**, accepted.
109. I. Muševič, M. Škarabot, D. Babic, N. Osterman, I. Poberaj, V. Nazarenko, and A. Nych (2004) *Phys Rev Lett* **93**, 187801.

110. M. Škarabot, M. Ravnik, D. Babic, N. Osterman, I. Poberaj, S. Žumer, I. Muševič, A. Nych, U. Ognysta, and V. Nazarenko (2006) *Physical Review E* **73**, 021705.
111. B. Lev, A. Nych, U. Ognysta, S. B. Chernyshuk, V. Nazarenko, M. Škarabot, I. Poberaj, D. Babic, N. Osterman, and I. Muševič (2006) *European Physical Journal E* **20**, 215-219.
112. S. Samitsu, Y. Takanishi, and J. Yamamoto (2010) *Nat Mater* **9**, 816-820.
113. M. Škarabot, Ž. Lokar, and I. Muševič (2013) *Physical Review E* **87**, 062501.
114. S. A. Tatarkova, D. R. Burnham, A. K. Kirby, G. D. Love, and E. M. Terentjev (2007) *Phys Rev Lett* **98**, 157801.
115. Y. Lansac, M. A. Glaser, N. A. Clark, and O. D. Lavrentovich (1999) *Nature* **398**, 54-57.
116. B. Zalar, O. D. Lavrentovich, H. R. Zeng, and D. Finotello (2000) *Physical Review E* **62**, 2252-2262.
117. B. Zupancic, S. Diez-Berart, D. Finotello, O. D. Lavrentovich, and B. Zalar (2012) *Phys Rev Lett* **108**, 257801.
118. T. A. Krentsel, O. D. Lavrentovich, and S. Kumar (1997) *Mol Cryst Liq Crys A* **304**, 463-469.
119. W. R. Folks, S. Keast, T. A. Krentzel, B. Zalar, H. Zeng, Y. A. Reznikov, M. Neubert, S. Kumar, D. Finotello, and O. D. Lavrentovich (1998) *Mol Cryst Liq Crys A* **320**, 77-88.
120. E. Grelet, M. P. Lettinga, M. Bier, R. Van Roij, and P. Van Der Schoot (2008) *J Phys-Condens Mat* **20**, 494213.
121. A. Patti, D. El Masri, R. Van Roij, and M. Dijkstra (2009) *Phys Rev Lett* **103**, 248304.
122. B. Mukherjee, C. Peter, and K. Kremer (2013) *Physical Review E* **88**, 010502(R).
123. O. P. Pishnyak, S. V. Shiyanovskii, and O. D. Lavrentovich (2011) *Phys Rev Lett* **106**, 047801.
124. O. P. Pishnyak, S. V. Shiyanovskii, and O. D. Lavrentovich (2011) *Journal of Molecular Liquids* **164**, 132-142.
125. H. A. Pohl Dielectrophoresis (Cambridge University Press, Cambridge 1978) p 580.
126. A. B. Golovin and O. D. Lavrentovich (2009) *Appl Phys Lett* **95**, 254104.
127. A. B. Golovin, J. Xiang, H. S. Park, L. Tortora, Y. A. Nastishin, S. V. Shiyanovskii, and O. D. Lavrentovich (2011) *Materials* **4**, 390-416.
128. G. Mchale, C. V. Brown, M. I. Newton, G. G. Wells, and N. Sampara (2011) *Phys Rev Lett* **107**, 186101.
129. V. H. Bodnar, Y. Kim, B. Taheri, and J. L. West (1999) *Mol Cryst Liq Crys A* **329**, 1017-1023.
130. H. W. Ren, S. T. Wu, and Y. H. Lin (2008) *Phys Rev Lett* **100**, 117801.
131. L. H. Hsu, K. Y. Lo, S. A. Huang, C. Y. Huang, and C. S. Yang (2008) *Appl Phys Lett* **92**, 181112.
132. D. W. Berreman (1975) *Journal of Applied Physics* **46**, 3746-3751.
133. C. Z. Vandoorn (1975) *Journal of Applied Physics* **46**, 3738-3745.
134. Z. Zou and N. Clark (1995) *Phys Rev Lett* **75**, 1799-1802.
135. Y. Mieda and K. Furutani (2005) *Appl Phys Lett* **86**, 101901.
136. M. Z. Bazant and O. I. Vinogradova (2008) *Journal of Fluid Mechanics* **613**, 125-134.
137. J. Fukuda, H. Stark, M. Yoneya, and H. Yokoyama (2004) *J Phys-Condens Mat* **16**, S1957-S1968.
138. T. Araki and H. Tanaka (2006) *J Phys-Condens Mat* **18**, L193-L203.
139. S. Khullar, C. F. Zhou, and J. J. Feng (2007) *Phys Rev Lett* **99**, 237802.
140. S. C. Glotzer and M. J. Solomon (2007) *Nat Mater* **6**, 557-562.
141. G. Helgesen, A. T. Skjeltorp, P. M. Mors, R. Botet, and R. Jullien (1988) *Phys Rev Lett* **61**, 1736-1739.
142. M. J. Stevens and G. S. Grest (1995) *Physical Review E* **51**, 5962-5975.
143. D. Zerrouki, J. Baudry, D. Pine, P. Chaikin, and J. Bibette (2008) *Nature* **455**, 380-382.
144. K. Nakahama, H. Kawaguchi, and K. Fujimoto (2000) *Langmuir* **16**, 7882-7886.
145. O. Cayre, V. N. Paunov, and O. D. Velev (2003) *Chem Commun*, 2296-2297.
146. S. Jiang, Q. Chen, M. Tripathy, E. Luijten, K. S. Schweizer, and S. Granick (2010) *Adv Mater* **22**, 1060-1071.
147. S. Fraden, A. J. Hurd, and R. B. Meyer (1989) *Phys Rev Lett* **63**, 2373-2376.
148. U. Dassanayake, S. Fraden, and A. Van Blaaderen (2000) *Journal of Chemical Physics* **112**, 3851-3858.
149. A. Yethiraj and A. Van Blaaderen (2003) *Nature* **421**, 513-517.
150. A. K. Agarwal and A. Yethiraj (2009) *Phys Rev Lett* **102**, 198301.
151. N. Li, H. D. Newman, M. Valera, I. Saika-Voivod, and A. Yethiraj (2010) *Soft Matter* **6**, 876-880.
152. J. H. E. Promislow, A. P. Gast, and M. Fermigier (1995) *Journal of Chemical Physics* **102**, 5492-5498.
153. F. M. Pedrero, M. Tirado-Miranda, A. Schmitt, and J. C. Fernandez (2006) *Journal of Chemical Physics* **125**, 084706.
154. S. Sacanna, D. J. Pine, and G.-R. Yi (2013) *Soft Matter* **9**, accepted.
155. S. B. Chernyshuk and B. I. Lev (2010) *Physical Review E* **81**, 041701.
156. T. Araki and H. Tanaka (2006) *J Phys-Condens Mat* **18**, L305-L314.
157. S. V. Shiyanovskii and O. V. Kuksenok (1998) *Mol Cryst Liq Crys A* **321**, 45-56.
158. B. T. Gettelfinger, J. A. Moreno-Razo, G. M. Koenig, J. P. Hernandez-Ortiz, N. L. Abbott, and J. J. De Pablo (2010) *Soft Matter* **6**, 896-901.
159. U. M. Ognysta, A. B. Nych, V. A. Uzunova, V. M. Pergamenschik, V. G. Nazarenko, M. Skarabot, and I. Musevic (2011) *Physical Review E* **83**, 041709.
160. I. Muševič, M. Škarabot, and M. Humar (2011) *J Phys-Condens Mat* **23**, 284112.
161. M. A. Gharbi, M. Nobili, M. In, G. Prevot, P. Galatola, J. B. Fournier, and C. Blanc (2011) *Soft Matter* **7**, 1467-1471.
162. Y. H. Kim, D. K. Yoon, M. C. Choi, H. S. Jeon, M. W. Kim, O. D. Lavrentovich, and H. T. Jung (2009) *Langmuir* **25**, 1685-1691.
163. J. L. West, A. Glushchenko, G. X. Liao, Y. Reznikov, D. Andrienko, and M. P. Allen (2002) *Physical Review E* **66**, 012702.

164. S. Kurihara, K. Ohta, T. Oda, R. Izumi, Y. Kuwahara, T. Ogata, and S.-N. Kim (2013) *Scientific Reports* **3**, 2167.
165. S. K. Smoukov, S. Gangwal, M. Marquez, and O. D. Velev (2009) *Soft Matter* **5**, 1285-1292.
166. Y.-K. Kim, B. Senyuk, and O. D. Lavrentovich (2012) *Nature Communications* **3**, 1133-1139.
167. Y.-K. Kim, M. Majumdar, B. I. Senyuk, L. Tortora, J. Seltmann, M. Lehmann, A. Jákli, J. T. Gleeson, O. D. Lavrentovich, and S. Sprunt (2012) *Soft Matter* **8**, 8880-8890.
168. A. Sengupta, U. Tkalec, and C. Bahr (2011) *Soft Matter* **7**, 6542-6549.
169. J. G. Cuennet, A. E. Vasdekis, L. De Sio, and D. Psaltis (2011) *Nat Photonics* **5**, 234-238.
170. A. Sengupta, B. Schulz, E. Ouskova, and C. Bahr (2012) *Microfluidics and Nanofluidics* **13**, 941-955.
171. A. Sengupta, C. Pieper, J. Enderlein, C. Bahr, and S. Herminghaus (2013) *Soft Matter* **9**, 1937-1946.
172. A. Sengupta, U. Tkalec, M. Ravnik, J. M. Yeomans, C. Bahr, and S. Herminghaus (2013) *Phys Rev Lett* **110**, 048303.
173. J. G. Guennet, A. E. Vasdekis, and D. Psaltis (2013) *Lab on a Chip* **13**, 2721-2726.
174. H. Morgan and N. G. Green AC Electrokinetics: colloids and nanoparticles (Research Studies Press, Baldock, UK 2003) p 324.
175. M. Z. Bazant, M. S. Kilic, B. D. Storey, and A. Ajdari (2009) *Advances in Colloid and Interface Science* **152**, 48-88.
176. M. Z. Bazant, M. S. Kilic, B. D. Storey, and A. Ajdari (2009) *New Journal of Physics* **11**, 075016.
177. C. L. Zhao and C. Yang (2012) *Microfluidics and Nanofluidics* **13**, 179-203.
178. V. A. Murtsovkin and G. I. Mantrov (1990) *Colloid Journal of the USSR* **52**, 933-936.
179. V. A. Murtsovkin and G. I. Mantrov (1991) *Colloid Journal of the USSR* **53**, 240-244.
180. N. I. Gamayunov, G. I. Mantrov, and V. A. Murtsovkin (1992) *Colloid Journal of the USSR* **54**, 20-23.
181. V. A. Murtsovkin and G. I. Mantrov (1992) *Colloid Journal of the USSR* **54**, 83-86.
182. M. Z. Bazant and T. M. Squires (2004) *Phys Rev Lett* **92**, 066101.
183. T. M. Squires and M. Z. Bazant (2004) *Journal of Fluid Mechanics* **509**, 217-252.
184. J. A. Levitan, S. Devasenathipathy, V. Studer, Y. X. Ben, T. Thorsen, T. M. Squires, and M. Z. Bazant (2005) *Colloids and Surfaces a-Physicochemical and Engineering Aspects* **267**, 122-132.
185. T. M. Squires and M. Z. Bazant (2006) *Journal of Fluid Mechanics* **560**, 65-101.
186. M. Z. Bazant and T. M. Squires (2010) *Current Opinion in Colloid & Interface Science* **15**, 203-213.
187. S. Gangwal, O. J. Cayre, M. Z. Bazant, and O. D. Velev (2008) *Phys Rev Lett* **100**, 058302.
188. M. Yoneya, J. I. Fukuda, H. Yokoyama, and H. Stark (2005) *Mol Cryst Liq Cryst* **435**, 735-745.
189. J. L. Billeter and R. A. Pelcovits (2000) *Physical Review E* **62**, 711-717.
190. J. B. Rovner, C. P. Lapointe, D. H. Reich, and R. L. Leheny (2010) *Phys Rev Lett* **105**, 228301.
191. I. Dierking, P. Cass, K. Syres, R. Cresswell, and S. Morton (2007) *Physical Review E* **76**, 021707.
192. A. K. Srivastava, M. Kim, S. M. Kim, M. K. Kim, K. Lee, Y. H. Lee, M. H. Lee, and S. H. Lee (2009) *Physical Review E* **80**, 051702.
193. I. Dierking, G. Biddulph, and K. Matthews (2006) *Physical Review E* **73**, 011702.
194. A. V. Ryzhkova, F. V. Podgornov, and W. Haase (2010) *Appl Phys Lett* **96**, 151901.
195. D. Sikharulidze (2005) *Appl Phys Lett* **86**, 033507.
196. S. Klein (2013) *Liquid Crystals Reviews* **1**, 52-82.
197. S. Hernàndez-Navarro, P. Tierno, J. Ignés-Mullol, and F. Sagués (2013) *Soft Matter* **9**, 7999-8004.
198. T. B. Jones Electromechanics of Particles (Cambridge University Press, Cambridge 1995) p 266.
199. G. Liao, I. I. Smalyukh, J. R. Kelly, O. D. Lavrentovich, and A. Jákli (2005) *Physical Review E* **72**, 031704.
200. J. Dobnikar, A. Snezhko, and A. Yethiraj (2013) *Soft Matter* **9**, 3693-3704.
201. A. G. Petrov The Lyotropic State of Matter (Gordon and Breach Science Publishers, Amsterdam, The Netherlands 1999).
202. J. Lydon (2011) *Liq Cryst* **38**, 1663-1681.
203. H. S. Park and O. D. Lavrentovich (2012) Lyotropic Chromonic Liquid Crystals: Emerging Applications *Liquid Crystals Beyond Displays: Chemistry, Physics, and Applications*, ed Q. Li (John Wiley & Sons, Hoboken, New Jersey), pp 449-484.
204. R. B. Meyer (1982) Macroscopic phenomena in nematic polymers. *Polymer Liquid Crystals*, ed W. R. K. A. Ciferri, R.B. Meyer (Academic Press, New York), pp 133-165.